\documentclass[12pt]{article}
\usepackage[usenames]{color}
\usepackage{setspace}
\usepackage{graphicx}
\usepackage{amsmath}
\usepackage{amssymb}
\usepackage{amsthm}
\usepackage{multirow}
\numberwithin{equation}{section}

\newcommand{\del}{\partial}

\newcommand{\bequ}{\begin{equation}}
\newcommand{\eequ}{\end{equation}}
\newcommand{\beqn}{\begin{eqnarray}}
\newcommand{\eeqn}{\end{eqnarray}}
\newcommand{\bctr}{\begin{center}}
\newcommand{\ectr}{\end{center}}
\newcommand{\bit}{\begin{itemize}}
\newcommand{\eit}{\end{itemize}}

\newcommand{\half}{{\frac12}}
\def\e{{\textrm e}}
\def\del{\partial}
\def\half{{\frac12}}

\def\vev#1{\langle #1 \rangle}
\def\del{\partial}
\def\half{{\frac12}}

\def\vev#1{\langle #1 \rangle}
\def\ket#1{|{#1}\rangle}
\def\bra#1{\langle {#1}|}
\def\del{\partial}
\def\dslash{\del\kern-0.55em\raise 0.14ex\hbox{/}}

\def\rough#1{\raise.3ex\hbox{$#1$\kern-.75em\lower1ex\hbox{$\sim$}}}

\newcommand{\PRD}[3]{{\it Phys. Rev.} {\bf D{#1}} (19{#3}) {#2}}

\newcommand{\PRDM}[3]{{\it Phys. Rev.} {\bf D{#1}} {#2} (20{#3})}

\newcommand{\NPB}[3]{{\it Nucl. Phys.} {\bf B{#1}} (19{#2}) {#3}}
\newcommand{\NPBM}[3]{{\it Nucl. Phys.} {\bf B{#1}} (20{#2}) {#3} }
\newcommand{\PLB}[3]{{\it Phys. Lett.} {\bf B{#1}} (19{#2}) {#3}}
\newcommand{\PLBM}[3]{{\it Phys. Lett.} {\bf B{#1}}, {#2} (20{#3})}

\newcommand{\ANN}[3]{{\it Ann. Phys. (N.Y.)} {\bf {#1}}, {#2} (19{#3})}

\begin{document}
\begin{flushright}
{\small KOBE-TH-16-09}\\%
\end{flushright}
\begin{center}
{\LARGE\bf 
Polyakov Loop \\ 
in Non-covariant Operator Formalism\\
}
\vskip 1.4cm
Makoto Sakamoto$^{(a)}$
\footnote{E-mail: dragon@kobe-u.ac.jp} and
Kazunori Takenaga$^{(b)}$
\footnote{E-mail: takenaga@kumamoto-hsu.ac.jp}
\\
\vskip 1.0cm
${}^{(a)}$ {\it Department of Physics, Kobe University, 
Rokkodai Nada, Kobe, 657-8501 Japan}
\\[0.2cm]
${}^{(b)}$ {\it Faculty of Health Science, Kumamoto
Health Science University, Izumi-machi, Kita-ku, Kumamoto 861-5598, Japan}
\\
\vskip 1.5cm
\begin{abstract}
We discuss a Polyakov loop in non-covariant operator formalism which consists 
of only physical degrees of freedom at finite temperature. It is pointed out that
although the Polyakov loop is expressed by a Euclidean time component
of gauge fields in a covariant path integral formalism, there is no direct counterpart
of the Polyakov loop operator in the operator formalism because the Euclidean time component 
of gauge fields is not a physical degree of freedom. We show that by starting with an operator
which is constructed in terms of only physical operators in the non-covariant operator formalism, the
vacuum expectation value of the operator calculated by trace formula can be rewritten 
into a familiar form of an expectation value of Polyakov loop in a covariant path integral
formalism at finite temperature for the cases of axial and Coulomb gauge.  
\end{abstract}
\end{center}
\vskip 1.0 cm
\newpage
%
\section{Introduction}
Gauge invariance is undoubtedly one of fundamental principles 
in particle physics.
Non-local gauge invariant operators as well as local ones play 
an important role in gauge theory.
An example of such non-local operators is a Wilson loop, 
which will be given by a line integral along a rectangular contour
that one side is taken to be in a space-like direction and 
another side to be in a time direction.
A non-trivial expectation value of the Wilson loop operator
will provide a signal of \lq\lq quark\rq\rq \,confinement 
in non-abelian gauge theories \cite{WilsonLoop}.

Another example of non-local gauge invariant operators is a 
Polyakov loop, which is similar to a Wilson loop 
but is given by a line integral along a Euclidean time axis.
In this paper, we will focus on the Polyakov loop operator 
whose explicit form is given by
$\textrm{tr}\,\mathcal{P}\exp\big( ig\int_{0}^{\beta} \!d\tau A_{\tau}\big)$,
where $\mathcal{P}$ denotes a path-ordered symbol,
$\beta$ is an inverse temperature and $A_{\tau}$ is a Euclidean time
component of gauge fields.
The Polyakov loop is known as an order parameter of the 
confinement-deconfinement phase transition at finite temperature \cite{PolyakovLoop1,PolyakovLoop2}.
Furthermore, it could provide an order parameter of gauge
symmetry breakings \cite{PolyakovLoop1, SymmetryBreaking, weiss, laine, fara}.

In order to confirm that the Polyakov loop is 
a physical observable at finite temperature, one might verify
that a zero mode of the Euclidean time component $A_{\tau}$
in the Polyakov loop cannot be eliminated by gauge transformations 
due to the periodicity with respect to the Euclidean time at finite temperature.
However, the statement that the Polyakov loop is physical
seems to be less obvious than we thought.\footnote{%
One might notice that Wilson lines along non-contractible loops
in extra dimensions \cite{hosotani} resemble a Polyakov loop along the Euclidean
time axis.
There seems formally no difference between them in a Euclidean 
path integral formulation.
However, there is an important difference.
A Polyakov loop becomes meaningless at zero temperature 
but Wilson lines do not. This is because the Euclidean time component
of gauge fields can be removed completely by gauge transformations
at zero temperature but the extra dimensional components cannot be.
}

A reason why the above statement is not so obvious is that
the Minkowski time component $A_{0}$ of gauge fields\footnote{%
In this paper, we will use $A_{\tau}$ and $A_{0}$ for the Euclidean
and Minkowski time components of gauge fields, respectively.
} 
is not a dynamical degree of freedom in gauge theory.
The $A_{0}$ component can be removed in an operator formalism,
and then the $A_{0}$ degree of freedom disappears completely from 
the Hilbert space of the theory, as explained below in some detail.
One might stress that the $A_{0}$ (or $A_{\tau}$) component is
included in a covariant path integral representation of gauge theory.
However, the $A_{0}$ (or $A_{\tau}$) degree of freedom turns out to be
introduced as \textit{an auxiliary field} in a path integral formalism
(see Section 3).

Quantization of gauge theory cannot be performed in a straightforward
fashion because it contains redundant gauge degrees of freedom 
due to gauge invariance.
Several ways to quantize it have been known and are expected
to be physically equivalent.\footnote{
We will not discuss the Gribov problem \cite{Gribov} in this paper.
The problem is expected not to be related directly to our issue.
}

One of reliable methods to quantize gauge theory is to remove
all unphysical degrees of freedom by explicitly imposing
a gauge fixing condition \cite{GaugeFixing}.
If the theory includes only physical degrees of freedom,
the quantization of the theory is rather straightforward,
though the form of the Hamiltonian would be messy.

For instance, in the axial gauge $A_{3}=0$ \cite{AxialGauge}, which will be discussed 
in this paper, the theory can be described only by the physical
degrees of freedom $A_{1}$ and $A_{2}$ for the gauge fields
because $A_{3}$ is taken to be zero by the axial gauge condition
and $A_{0}$ is also removed by use of the equation of motion
for $A_{0}$, which should be regarded as a constraint equation
since it includes no time derivative of $A_{0}$.\footnote{%
If we take the temporal gauge $A_{0}=0$ \cite{TemporalGauge}, 
$A_{0}$ can be eliminated rather directly.
}
Therefore, the gauge part of the Hilbert space in the axial
gauge formalism can be spanned by the states $\{\,\vert\,A_{1} , A_{2}\,\rangle\,\}$,
and any physical operators have to be constructed in terms of
$A_{k}\ (k=1,2)$ and their conjugate momenta $\Pi_{k}$
(and also fermion/scalar fields).

Now we can raise an issue of how to construct a Polyakov 
loop operator in the axial gauge operator formalism.
Since the time component is completely eliminated from the
spectrum in the operator formalism, it seems non-trivial to construct
a Polyakov loop operator without $A_{0}$ or $A_{\tau}$.

One of our main purposes is to present an explicit operator form
corresponding to a Polyakov loop operator, in the operator formalism
which contains only physical degrees of freedom: $A_{k}, \Pi_{k}\ (k=1,2)$ 
and matter fields.
Another purpose of this paper is to show that an expectation value
of the above operator given in the non-covariant operator
formalism can be rewritten into a familiar form of an expectation
value of a Polyakov loop in a covariant path integral formalism
at finite temperature.
In the rewriting, we will see where the Euclidean time component 
$A_{\tau}$ comes from and also clarify how the operator given by physical
operators turns to the standard form of the
Polyakov loop given by $A_{\tau}$.

This paper is organized as follows.
In the next section, we take the axial gauge to quantize a non-abelian gauge
theory, and remove all the unphysical degrees of freedom from a Lagrangian.
Then, we derive a Hamiltonian consisting of only physical ones.
In Section 3, we first construct a finite temperature path integral 
representation in the axial gauge by use of the trace formula \cite{finitet}, 
and then rewrite it into a covariant path integral form with all the
gauge degrees of freedom. We will see that the Euclidean time component of the gauge fields is
recovered as a Gaussian auxiliary field.
Our main results will be given in Section 4.
We propose an operator that is expected to represent a Polyakov loop
operator in the axial gauge and is written in terms of physical
operators only.
We show that the operator turns to the standard form of the Polyakov
loop in a covariant Euclidean path integral formula at finite temperature.
In Section 5, we reformulate the analysis in the Coulomb gauge.
Section 6 is devoted to conclusions.
Some details of calculations will be given in Appendices.
\section{Hamiltonian in Axial Gauge}
In this section, we present a canonical formulation of non-abelian gauge theory coupled with fermion in axial gauge. 
Although the results given in Sections $2$ and $3$ will not be new \cite{GaugeFixing}, we shall explain them to
make our discussions in Section $4$ comprehensible. We 
consider an $SU(N)$ gauge theory whose Lagrangian is given by
\begin{equation}
{\cal L} =\frac{-1}{4}F^a_{\mu\nu}F^{a\mu\nu} + \bar\psi i\gamma^{\mu}D_{\mu}\psi,
\label{eq:lag}
\end{equation} 
where 
\begin{equation}
F^a_{\mu\nu}\equiv \del_{\mu}A^a_{\nu} - \del_{\nu}A^a_{\mu} + gf^{abc}A^b_{\mu}A^c_{\nu},\quad 
D_{\mu}\psi \equiv \left(\del_{\mu} -ig A_{\mu}^aT^a \right)\psi .
\end{equation}
The representation of the fermion $\psi$ under the $SU(N)$ gauge group can be arbitrary, but for simplicity, we 
take it to be the fundamental representation. Throughout this paper, we use the convention
that repeated indices are summed over (otherwise stated) 
and $\mu, \nu,\cdots = 0, 1, 2, 3~(\mbox{or}~~\tau, 1, 2, 3),~i, j=1, 2, 3,~k, l, \cdots =1, 2,~a, b,\cdots =1, 2, \cdots, N^2-1$.
The equation of motion for the gauge field $A^a_{\mu}$ followed from the Lagrangian (\ref{eq:lag}) is
\begin{equation}
(D_{\mu})^{ab}F^{b\mu\nu} + J_F^{a\nu} = 0,
\label{eq:eom}
\end{equation}
where $J_F^{a\nu} \equiv  g\bar\psi \gamma^{\nu}T^a \psi$ and
$(D_{\mu})^{ab} = \delta^{ab}\del_{\mu} + gf^{acb}A_{\mu}^c$. We 
impose the axial gauge 
\begin{equation}
A_3^a=0.
\label{eq:axialgauge}
\end{equation}
Then, we can formally solve the $\nu =0$ component of the equation of motion 
for $A_0^a$ and obtain that 
\begin{equation}
A_0^a= ({\cal D}^{-1})^{ab}\Bigl((D_k\dot{A}_k)^b + J_F^{b0}\Bigr)\qquad (k=1,2 ).
\label{eq:constraint}
\end{equation}
Here the operator  $({\cal D}^{-1})^{ab}$ is formally defined as the inverse 
of $({\cal D})^{ab} \equiv (D_k^2)^{ab} + \delta^{ab}(\del_3)^2$. 
%
%

We define the canonical momentum for the gauge field $A_k^a (k=1, 2)$ by
\begin{equation}
\Pi_A^{ak}\equiv\frac{\del {\cal L}(A_k, \psi)_{\rm axial}}{\del \dot{A}_k^a},
\end{equation}
where  ${{\cal L}}(A_k, \psi)_{\rm axial}$ is obtained by imposing the axial gauge on 
the Lagrangian (\ref{eq:lag}) and eliminating $A_0^a$ in the Lagrangian (\ref{eq:lag}) by 
using the equation (\ref{eq:constraint}), so that it is written in terms of only physical
degrees of freedom. The explicit form of 
${{\cal L}}(A_k, \psi)_{\rm axial}$ is given by Eq.(\ref{eq:tildelagaxial}) in the 
Appendix \ref{sec:canonicalaxial}, from which the canonical momentum for $A_k^a$ is calculated as
\begin{equation}
\Pi_A^{ak}=(M)_{kl}^{ab}\dot{A}_l^b - (D_k{\cal D}^{-1}J_F^0)^a\qquad (l=1,2),
\label{eq:momentum}
\end{equation}
where 
\begin{equation}
(M)_{kl}^{ab} = \delta^{ab}\delta_{kl} - (D_k{\cal D}^{-1}D_l)^{ab} .
\label{eq:mexplicit}
\end{equation}
By solving the equation (\ref{eq:momentum}) for $\dot{A}_k^a$, we have
\begin{equation}
\dot{A}_k^a = (M^{-1})_{kl}^{ab}\Bigl(\Pi_A^{bl} + (D_l{\cal D}^{-1}J_F^0)^b\Bigr),
\label{eq:dotakaxial}
\end{equation}
where $(M^{-1})_{kl}^{ab}$ is defined as the inverse of $(M)_{kl}^{ab}$, whose explicit form is 
\begin{equation}
(M^{-1})_{kl}^{ab}=\delta^{ab}\delta_{kl} + (D_k(\del_3)^{-2}D_l)^{ab}.
\label{eq:minv}
\end{equation}
It is straightforward to check $(MM^{-1})^{ab}_{km}=\delta^{ab}\delta_{km}$ by 
using $({\cal D})^{ab}=(D_k^2)^{ab}+\delta^{ab}(\del_3)^2$. As for 
the fermion $\psi$, we take the right derivative with respect to the Grassmann variable, and the canonical 
momentum for $\psi$ is
\begin{equation}
\Pi_{\psi}\equiv\frac{\del {{\cal L}}(A_k, \psi)_{\rm axial}}{\del\dot{\psi}} ={\bar\psi}i\gamma^0 .
\label{eq:momentumf}
\end{equation}

Then, the Hamiltonian is 
\begin{eqnarray}
{\cal H}_{\rm axial} &\equiv & \Pi_A^{ak}\dot{A}_k^a +\Pi_{\psi}\dot{\psi} - {{\cal L}}(A_k, \psi)_{\rm axial} 
\nonumber\\
&=&
\half \Pi_A^{ak}(M^{-1}_{kl}\Pi_A^l)^a + \Pi_A^{ak}(M^{-1}_{kl}D_l{\cal D}^{-1}J_F^0)^a
\nonumber\\
&&+\half (D_k{\cal D}^{-1}J_F^0)^a(M^{-1}_{kl}D_l{\cal D}^{-1}J_F^0)^a -\half J_F^{a0}({\cal D}^{-1}J_F^0)^a
\nonumber\\
&&+\frac{1}{4}F_{kl}^a F_{kl}^a +\half \del_3A_k^a\del_3A_k^a - \bar\psi i \gamma^k D_k \psi 
-\bar\psi i \gamma^3\del_3\psi .
\label{eq:hamiltonian}
\end{eqnarray}
In deriving the Hamiltonian, we have performed the partial integration with 
resect to $(M^{-1})_{kl}^{ab}$. The Hamiltonian is written in terms of only physical degrees of 
freedom $A_k^a, \Pi_A^{ak}, \psi$ and $\Pi_{\psi}$ as it should, though it is messy, including the
non-local terms. It may be appropriate to mention here that one is able to rewrite (\ref{eq:hamiltonian}) 
into another form presented by Eq.(\ref{eq:anoham}) in the Appendix \ref{sec:canonicalaxial}, in 
which the self-energy of the color charge densities of fermion manifestly 
appears. We shall use the Hamiltonian (\ref{eq:hamiltonian}) in the discussions given below.
\section{Trace Formula in Axial Gauge at Finite Temperature}
Let us first define the vacuum expectation value of an
operator $\hat{\cal O}$ at finite temperature by the trace formula,
\begin{equation}
\vev{\hat{\cal O}}_{\beta}\equiv {\rm Tr}\left(\hat{{\cal O}}~\e^{-\beta {\hat H}}\right)
\quad\mbox{with}\quad{\hat H}\equiv \int d^3{\bf x}~\hat{\cal H},
\label{eq:trace}
\end{equation}
where $\beta$ stands for the inverse of the temperature $T$. If one chooses a non-covariant 
gauge such as the axial or Coulomb gauge, the 
operator $\hat{{\cal O}}$ and the Hamiltonian ${\hat H}$ are written in terms of only physical degrees 
of freedom, {\it i.e.} ${\hat A}_k^a, \hat\psi$ and their canonical momenta, ${\hat{\Pi}}_A^{ak}, {\hat\Pi}_{\psi}$, so that the 
trace in Eq.(\ref{eq:trace}) must be taken over the physical states alone, denoted here
by $A_{\rm phys}=\{A_k^a, \psi\}$,
\begin{eqnarray}
&&
{\rm Tr}\left({\cal O}({\hat\Pi}_A^{ak}, {\hat A}_k^a, {\hat\Pi}_{\psi}, {\hat\psi})~
\e^{-\beta H( {\hat\Pi}_A^{k}, {\hat A}_k, {\hat\Pi}_{\psi}, {\hat\psi})}\right)_{\rm gauge} 
\nonumber\\
&&=\int {\cal D}A_{\rm phys}\bra{A_{\rm phys}}
{\cal O}( {\hat\Pi}_A^{ak}, {\hat A}_k^a, ,{\hat\Pi}_{\psi}, {\hat\psi})
~\e^{-\beta H( {\hat\Pi}_A^{k}, {\hat A}_k, {\hat\Pi}_{\psi}, {\hat\psi})}\ket{A_{\rm phys}}_{\rm gauge}.
\label{eq:tracegauge}
\end{eqnarray}
It should be understood that bosons (fermions) must obey 
the (anti) periodic boundary condition for the Euclidean time direction
because of the quantum statistics at finite temperature.

Our aim is to find the operator ${\cal O}({\hat\Pi}_A^{ak}, {\hat A}_k^a, {\hat\Pi}_{\psi}, {\hat\psi})$
which satisfies 
\begin{eqnarray}
&&
{\rm Tr}\left({\cal O}({\hat A}_k^a, {\hat\psi}, {\hat\Pi}_A^{ak},{\hat\Pi}_{\psi})
\e^{-\beta H({\hat A}_k^a, {\hat\psi}, {\hat\Pi}_A^{ak},{\hat\Pi}_{\psi})}\right)_{\rm gauge} 
=\int{\cal D}A_{\mu}{\cal D}{\bar\psi}{\cal D}\psi~
{\rm det}(M_{\rm FP}^{\rm gauge})\prod_{x,a}\delta(\chi^a_{\rm gauge}(x))\nonumber\\
&&\times~ 
{\rm tr}\Bigl({\cal P}~\e^{ig\int_0^{\beta}d\tau A_{\tau}(\tau,{\bf x}_0)}\Bigr)
~{\rm exp}
\int_0^{\beta}d\tau\int d^3{\bf x}
\Bigl\{
\frac{-1}{4}F^a_{\mu\nu}F^a_{\mu\nu}  - {\bar\psi}i \gamma_{\mu}D_{\mu}\psi
\Bigr\},
\label{eq:aimshiki}
\end{eqnarray}
where $M_{\rm FP}^{\rm gauge}$ and $\delta(\chi^a_{\rm gauge}(x))$ are the
Faddeev-Popov determinant and the gauge condition for the chosen gauge, respectively and $\mu, \nu =\tau, 1, 2, 3$.
We shall consider the cases of the axial and Coulomb gauge in this paper. 
The right-hand-side of Eq.(\ref{eq:aimshiki}) is nothing but the vacuum expectation value of the
Polyakov loop operator in the covariant path integral form under the chosen gauge condition.
It should be noticed that the left-hand-side of Eq.(\ref{eq:aimshiki}) is given by the trace formula 
in the non-covariant operator formalism by use of the physical degrees of freedom alone, not 
including $A_{\tau}$, especially that the operator ${\cal O}( {\hat\Pi}_A^{ak}, {\hat A}_k^a, {\hat\Pi}_{\psi}, {\hat\psi})$ 
has to be constructed by the physical operators only. Therefore, it seems to be a non-trivial
problem whether or not there exists such an operator ${\cal O}( {\hat\Pi}_A^{ak}, {\hat A}_k^a, {\hat\Pi}_{\psi}, {\hat\psi})$ 
satisfying the equation (\ref{eq:aimshiki}). Furthermore, we need to clarify where the unphysical degree
of freedom $A_{\tau}$ comes from in deriving the right-hand-side of Eq.(\ref{eq:aimshiki})
from the left-hand-side of it, and manage the path ordered product of the Polyakov loop in the
trace formula.

Before we construct the operator $\hat{\cal O}$ explained above, let us first study the partition function at 
finite temperature, say, ${\hat{\cal O}} =\hat{\bf 1}$ in the case of the axial gauge. 
In this case what we would like to show is that for the Hamiltonian (\ref{eq:hamiltonian}) the equation,
\begin{eqnarray}
&&{\rm Tr}\left(\e^{-\beta H({\hat \Pi}_A^{k}, {\hat A}_k, \hat{\Pi}_{\psi}, \hat{\psi})}\right)_{\rm axial} 
\nonumber\\
&&
=\int{\cal D}A_{\mu}{\cal D}{\bar\psi}{\cal D}\psi~
{\rm det}(\del_3)\prod_{x,a}\delta(A_3^a)~
{\rm exp} \int_0^{\beta}d\tau\int d^3{\bf x}
\Bigl\{
\frac{-1}{4}F^a_{\mu\nu}F^a_{\mu\nu}  - {\bar\psi}i \gamma_{\mu}D_{\mu}\psi
\Bigr\}
\label{eq:partitionaxial}
\end{eqnarray}
holds where the right-hand-side of Eq.(\ref{eq:partitionaxial}) is the standard path integral representation 
for the partition function at finite temperature in the axial gauge. 
This may be too pedagogical, but helpful for the later discussions. 
Using the completeness relation for the canonical momenta $\Pi_A^{ak}$ and $\Pi_{\psi}$, the 
left-hand-side of Eq.({\ref{eq:partitionaxial}) can be written in the path integral form as
\begin{eqnarray}
&&{\rm Tr}\left(\e^{-\beta H(\hat{\Pi}_A^k, \hat{A}_k, \hat{\Pi}_{\psi}, \hat{\psi})}\right)_{\rm axial}
\nonumber\\
&&=
\int {\cal D}\Pi_A^{ak}{\cal D}A_k^a{\cal D}\Pi_{\psi}{\cal D}\psi ~
{\rm exp}
\int_0^{\beta}d\tau \int d^3{\bf x}
\Bigl\{
i\Pi_{A}^{ak}\dot{A}_k^a + i\Pi_{\psi}\dot{\psi} -{\cal H}_{\rm axial}
\Bigr\},
\label{eq:partaxial}
\end{eqnarray}
where ${\cal H}_{\rm axial}$ is given by Eq. (\ref{eq:hamiltonian}) and $\dot{A}_k^a, \dot{\psi}$
stands for the derivative with respect to the Euclidean time
$\tau$, $\dot{A}_k^a=\del_{\tau}A_k^a, \dot{\psi}=\del_{\tau}\psi$. This notation will be used hereafter. 
Let us note that the imaginary unit  \lq\lq~{\it i}~\rq\rq~in front of $\Pi_A^{ak}\dot{A}_k^a$ and $\Pi_{\psi}\dot{\psi}$ 
does not come from the Euclideanization, but from taking the inner product of
the \lq\lq coordinate\rq\rq $\phi({\bf x})$ and the \lq\lq momentum\rq\rq $\Pi_{\phi}({\bf x})$, 
\begin{equation} 
\vev{\phi | \Pi_{\phi}} \propto {\rm exp}\Bigl\{ i\int d^3{\bf x}~ \Pi_{\phi}({\bf x})\phi({\bf x})\Bigr\}.
\end{equation}

The exponent of the right-hand-side of Eq.(\ref{eq:partaxial}) is quadratic with 
respect to $\Pi_A^{ak}$, whose explicit form is 
given by Eq.(\ref{eq:quadraaxial}) in the Appendix \ref{sec:canonicalaxial}, so that 
we can perform the Gaussian integration ${\cal D}\Pi_A^{ak}$. Then, we arrive at
\begin{eqnarray}
&&{\rm Tr}\left(\e^{-\beta H(\hat{\Pi}_A^k, \hat{A}_k, \hat{\Pi}_{\psi}, \hat{\psi})}\right)_{\rm axial}
=\int {\cal D}A_k^a{\cal D}\Pi_{\psi}{\cal D}\psi ~{\rm det}^{-\half}(M^{-1})
~{\rm exp}~\int_0^{\beta}d\tau\int d^3{\bf x} \nonumber\\
&&\times \Bigl\{
-\half \dot{A}_k^a\dot{A}_k^a -\half (D_k\dot{A}_k)^a({\cal D}^{-1}D_l\dot{A}_l)^a 
-i\dot{A}_k^a(D_k{\cal D}^{-1}J_F^0)^a 
+\half J_F^{a0}({\cal D}^{-1}J_F^0)^a \nonumber\\
&&~~
-\frac{1}{4}F_{kl}^a F_{kl}^a -\half \del_3A_k^a\del_3A_k^a 
+i\Pi_{\psi}\dot{\psi}
+ \bar\psi i \gamma^k D_k \psi  +\bar\psi i \gamma^3\del_3\psi 
\Bigr\}.
\label{eq:afterdpi}
\end{eqnarray}
In order to restore the $A_{\tau}$ degree of freedom, let us consider the Gaussian integral given by
\begin{eqnarray}
&&1=\int {\cal D}A_{\tau}^a ~{\rm det}^{\half}({\cal D})~{\rm exp}\int_0^{\beta}d\tau \int d^3{\bf x}
\Bigl\{-\half\Bigl(iA_{\tau}^a -({\cal D}^{-1})^{ac}\bigl(i (D_k\dot{A}_k)^c +J_F^{c0}   \bigr)\Bigr)\nonumber \\
&&~~\times
({\cal D})^{ad}\Bigl(iA_{\tau}^d -({\cal D}^{-1})^{de}\bigl(i (D_l\dot{A}_l)^e +J_F^{e0}   \bigr)\Bigr)\Bigr\}.
\label{eq:gaussatauaxial}
\end{eqnarray}
By inserting the equation (\ref{eq:gaussatauaxial}) into 
Eq.(\ref{eq:afterdpi}), some of the terms in the exponent in Eqs.(\ref{eq:afterdpi}) and (\ref{eq:gaussatauaxial}) 
are cancelled each other, as shown in the Appendix \ref{sec:traceaxial}, where we explicitly present
the expansion of the exponent in the Gaussian integral (\ref{eq:gaussatauaxial}) by Eq.(\ref{eq:expansionaxial}).
Then, we obtain that
 \begin{eqnarray}
&&{\rm Tr}\left(\e^{-\beta H(\hat{\Pi}_A^k, \hat{A}_k, \hat{\Pi}_{\psi}, \hat{\psi})}\right)_{\rm axial}
=\int {\cal D}A_{\tau}^a{\cal D}A_k^a{\cal D}\Pi_{\psi}{\cal D}\psi~{\rm det}^{-\half}(M^{-1})
 {\rm det}^{\half}({\cal D})\nonumber\\
 &&\times~ 
 {\rm exp}\int_0^{\beta}d\tau \int d^3{\bf x} 
 \Bigl\{
-\half \dot{A}_k^a\dot{A}_k^a - \frac{1}{4}F_{kl}^aF_{kl}^a -\half \del_3A_k^a \del_3 A_k^a \nonumber\\
&&
-\half  (D_k A_{\tau})^a(D_k A_{\tau})^a 
-\half \del_3A_{\tau}^a \del_3A_{\tau}^a +\dot{A}_k^a(D_kA_{\tau})^a
\nonumber\\
&&+i\Pi_{\psi}\dot{\psi}
+ \bar\psi i \gamma^k D_k \psi  +\bar\psi i \gamma^3\del_3\psi + iA_{\tau}^a J_F^{a0} 
\Bigr\}.
\label{eq:axialprefinal}
 \end{eqnarray}
Let us note that $A_{\tau}^a$ is recovered as the auxiliary field through the Gaussian integral (\ref{eq:gaussatauaxial}).
We finally introduce the factor,
\begin{equation}
1=\int {\cal D}A_3^a \prod_{x, a}\delta(A_3^a) 
\label{eq:a3delta}
\end{equation}
in order to restore the $A_3^a$ degree of freedom. By inserting it 
into Eq.(\ref{eq:axialprefinal}), the exponent in Eq.(\ref{eq:axialprefinal}) is summarized into the covariant form of the 
Lagrangian thanks to the delta function in Eq.(\ref{eq:a3delta}). And the determinants 
in Eq.(\ref{eq:axialprefinal}) are evaluated as
\begin{equation}
{\rm det}^{-\half}(M^{-1}) {\rm det}^{\half}({\cal D}) = {\rm det}(\del_3),
\label{eq:determinantaxial}
\end{equation}
which is shown in the Appendix \ref{sec:detaxial}. Hence, we finally obtain that 
\begin{eqnarray}
&&
{\rm Tr}\left(\e^{-\beta H(\hat{\Pi}_A^k, \hat{A}_k, \hat{\Pi}_{\psi}, \hat{\psi})}\right)_{\rm axial}
\nonumber\\
&=&
\int {\cal D}A_{\mu}^a {\cal D}{\bar\psi}{\cal D}\psi~{\rm det}(\del_3)
\prod_{x, a}\delta(A_3^a)~
{\rm exp} \int_0^{\beta}d\tau\int d^3{\bf x}
\Bigl\{
\frac{-1}{4}F^a_{\mu\nu}F^a_{\mu\nu}  - {\bar\psi}i \gamma_{\mu}D_{\mu}\psi
\Bigr\}.
\label{eq:partaxialfinal}
\end{eqnarray}
where we have used ${\cal D}\Pi_{\psi}={\cal D}\bar\psi$ and defined $\gamma_{\tau}\equiv -i\gamma^0$.
This is the equation (\ref{eq:partitionaxial}) that we would like to prove. 
\section{Polyakov Loop in Axial Gauge}
Let us discuss the Polyakov loop operator in the axial gauge which is one of our main purposes in
the paper. As explained in the introduction, the gauge part of the 
Hilbert space in the axial gauge does not contain the gauge field $A_{\tau}$, from which the Polyakov loop
operator is usually defined. Thus, the problem is how to construct the operator 
written in terms of only physical degrees of freedom in the axial gauge, 
corresponding to the Polyakov loop operator. Our aim is to present an explicit form of the 
operator ${\cal O}(\hat{\Pi}_A^{ak}, \hat{A}_k^a, \hat{\Pi}_{\psi}, \hat{\psi})$ which satisfies 
\begin{eqnarray}
&&
{\rm Tr}\left({\cal O}(\hat{\Pi}_A^{ak}, \hat{A}_k^a, \hat{\Pi}_{\psi}, \hat{\psi})
~\e^{-\beta H(\hat{\Pi}_A^k, \hat{A}_k, \hat{\Pi}_{\psi}, \hat{\psi})}\right)_{\rm axial}
\nonumber\\
&=&
\int {\cal D}A_{\mu}^a {\cal D}{\bar\psi}{\cal D}\psi~{\rm det}(\del_3)
\prod_{x, a}\delta(A_3^a)~
{\rm tr}\Bigl({\cal P}~{\rm exp}~\Bigl\{ig\int_0^{\beta}d\tau A_{\tau}(\tau,{\bf x}_0)\Bigr\}\Bigr) \nonumber\\
&&
\times~ {\rm exp}\int_0^{\beta}d\tau\int d^3{\bf x}
\Bigl\{
\frac{-1}{4}F^a_{\mu\nu}F^a_{\mu\nu}  - {\bar\psi}i \gamma_{\mu}D_{\mu}\psi
\Bigr\},
\label{eq:polyaxial}
\end{eqnarray}
where the operator ${\cal O}(\hat{\Pi}_A^{ak}, \hat{A}_k^a, \hat{\Pi}_{\psi}, \hat{\psi})$ must 
be written in terms of only physical degrees of freedom in the axial gauge. The trace of the left-hand-side 
is taken over the physical state alone in the axial gauge where there is no $A_{\tau}^a$ degree of freedom.
The right-hand-side is, of course, the vacuum expectation value of the 
Polyakov loop operator in the covariant path integral form with the axial gauge.

In our trials to find operators satisfying the equation (\ref{eq:polyaxial}), we have observed that 
the non-abelian nature of the path-ordered product of the Polyakov loop in Eq.(\ref{eq:polyaxial}) 
is an obstacle to rewrite the expectation value of the 
operator ${\cal O}(\hat{\Pi}_A^{ak}, \hat{A}_k^a, \hat{\Pi}_{\psi}, \hat{\psi})$ 
in the trace formula of the left-hand-side of Eq.(\ref{eq:polyaxial}) into the covariant path integral form in the
right-hand-side of Eq.(\ref{eq:polyaxial}). Thus, it seems to be necessary to transform
the right-hand-side of Eq.(\ref{eq:polyaxial}) into a path integral form without path-ordered product.

To this end, let us consider the gauge transformations such as
\begin{eqnarray}
A_{\tau}^{a}(\tau, {\bf x})T^{a} &=&U(\tau) \Bigl({A^{\prime}}_{\tau}^{a}(\tau, {\bf x})T^a
+\frac{i}{g}\del_{\tau}\Bigr)U^{\dagger}(\tau),
\label{eq:gtrf}\\
A_i^a(\tau, {\bf x})T^a& =&U(\tau) {A^{\prime}}_i^{a}(\tau, {\bf x})T^a U^{\dagger}(\tau),\label{eq:gtrf2}\\
\psi(\tau,{\bf x}) &=&U(\tau)\psi^{\prime}(\tau, {\bf x})\label{eq:gtrf3},
\end{eqnarray}
where the unitary matrix $U(\tau)$ is assumed to depend only on the Euclidean time $\tau$ 
with the periodic boundary condition $U(\tau +\beta)=U(\tau)$ in order not to contradict with
the quantum statistics at finite temperature. We notice that there is no inhomogeneous term
in the right-hand-side of Eq.(\ref{eq:gtrf2}) because $U(\tau)$ does not depend on the
spatial coordinate $\bf x$. If we formally write the delta 
functions $\displaystyle{\prod_{x, a}}\delta(A_3^a(\tau,{\bf x}))$ as 
$\displaystyle{\prod_{x}}\delta(A_3(\tau,{\bf x}))$ with $A_3=A_3^aT^a$, it will be
obvious that the delta functions are invariant under the gauge transformation
(\ref{eq:gtrf2}), {\it i.e.}
\begin{equation}
\prod_x \delta(A_3(\tau, {\bf x}))=
\prod_x \delta(A_3^{\prime}(\tau, {\bf x})).
\label{eq:gtrfa3}
\end{equation}

We choose the unitary matrix $U(\tau)$ to diagonalize the Polyakov loop
by the gauge transformation (\ref{eq:gtrf}) as follows: The path-ordered 
product ${\cal P}~\e^{ig\int_0^{\beta}d\tau A_{\tau}^a(\tau, {\bf x}_0)T^a}$ transforms 
under the gauge transformation (\ref{eq:gtrf}) as
\begin{eqnarray}
{\cal P}~\e^{ig\int_0^{\beta}d\tau A_{\tau}^a(\tau, {\bf x}_0) T^a }
&=&U(\beta)\Biggl[{\cal P}~\e^{ig\int_0^{\beta}d\tau {A^{\prime}}_{\tau}^a(\tau, {\bf x}_0) T^a }\Biggr] U^{\dagger}(0)
\nonumber\\
&=&U(0)\Biggl[{\cal P}~\e^{ig\int_0^{\beta}d\tau {A^{\prime}}_{\tau}^a(\tau, {\bf x}_0) T^a }\Biggr] U^{\dagger}(0).
\label{eq:ataudiag}
\end{eqnarray} 
Since ${\cal P}~\e^{ig\int_0^{\beta}d\tau {A^{\prime}}_{\tau}^a(\tau, {\bf x}_0)T^a}$ is a unitary 
matrix fixed at ${\bf x}_0$, it can be diagonalized by a unitary matrix $U(0)$. Then, we can rewrite the Polyakov loop
into the form
\begin{eqnarray}
{\rm tr}\Bigl({\cal P}~\e^{ig\int_0^{\beta}d\tau A_{\tau}^a(\tau,{\bf x}_0)T^a}\Bigr)
&=&{\rm tr}\Bigl(
\exp\Bigl\{g\int_0^{\beta}d\tau {A^{\prime}}_{\tau}^{\tilde a}(\tau,{\bf x}_0)T^{\tilde a}\Bigr\}
\Bigr)\nonumber\\
&=&\sum_{\alpha=1}^D 
\exp\Bigl\{ig\int_0^{\beta}d\tau {A^{\prime}}_{\tau}^{\tilde a}(\tau, {\bf x}_0)(T^{\tilde a})_{\alpha\alpha}\Bigr\},
\label{eq:atausum}
\end{eqnarray}
where $\{T^{\tilde a}, {\tilde a}=1, 2, \cdots, N-1\}$ are generators of the Cartan subalgebra of $SU(N)$ 
with the diagonal form
\begin{equation}
(T^{\tilde a})_{\alpha\beta}=(T^{\tilde a})_{\alpha\alpha}\delta_{\alpha\beta}\qquad (\alpha, \beta=1, 2, \cdots, D).
\end{equation}
Here, $D$ stands for the dimension of the representation of $\{T^a\}$. ($D=N$ for the fundamental representation.)
Thus, the Polyakov loop turns out to reduce to the one defined for the $U(1)$ subgroups in the $SU(N)$ gauge 
group. Thanks to the reduction, the trace in the original expression (\ref{eq:polyaxial}) is replaced by 
the summation with respect to the abelian part of the Polyakov loop in Eq.(\ref{eq:atausum})
(note that $(T^{\tilde a})_{\alpha\alpha}$ is just a number) and one does not need to take care of 
the path-ordered integral because of the abelian nature of the $U(1)$ gauge group.

It is convenient for transparent calculations to use the notation defined by
\begin{equation}
(\tilde{T}^a)_{\alpha\alpha} \equiv \left\{
\begin{array}{cl}
(T^{\tilde a})_{\alpha\alpha} &\mbox{for}~~a={\tilde a}\quad (\mbox{not~summed over}~\alpha),
\\
0 &\mbox{for}~~a\neq {\tilde a}.
\end{array}\right.
\label{eq:cartan}
\end{equation}
By taking account of the above discussions, the equation (\ref{eq:polyaxial}) becomes 
\begin{eqnarray}
&&{\rm Tr}\left({\cal O}(\hat{\Pi}_A^{ak}, \hat{A}_k^a, \hat{\Pi}_{\psi}, \hat{\psi})
~\e^{-\beta H(\hat{\Pi}_A^k, \hat{A}_k, \hat{\Pi}_{\psi}, \hat{\psi})}\right)_{\rm axial}
\nonumber\\
&=&\sum_{\alpha=1}^D 
\int {\cal D}A_{\mu}^a {\cal D}{\bar\psi}{\cal D}\psi~{\rm det}(\del_3)
\prod_{x, a}\delta(A_3^a)~
{\rm exp}~\Bigl\{ig \int_0^{\beta}d\tau A_\tau^{a}(\tau, {\bf x}_0)({\tilde T}^a)_{\alpha\alpha} \Bigr\}
\nonumber\\
&&
\times ~{\rm exp} \int_0^{\beta}d\tau\int d^3{\bf x}
\Bigl\{
\frac{-1}{4}F^a_{\mu\nu}F^a_{\mu\nu}  - {\bar\psi}i \gamma_{\mu}D_{\mu}\psi
\Bigr\},
\label{eq:polyaxial2}
\end{eqnarray}
where we have removed the dash $\prime$ from all the fields. Thus, our aim is now to find the operator
${\cal O}(\hat{\Pi}_A^{ak}, \hat{A}_k^a, \hat{\Pi}_{\psi}, \hat{\psi})$ which satisfies the above relation.

We propose the operator ${\cal O}(\hat{\Pi}_A^{ak}, \hat{A}_k^a, \hat{\Pi}_{\psi}, \hat{\psi})$
 satisfying the Eq.(\ref{eq:polyaxial2}) as
\begin{equation}
{\cal O}(\hat{\Pi}_A^{ak}, \hat{A}_k^a, \hat{\Pi}_{\psi}, \hat{\psi})
=\sum_{\alpha=1}^D
{\rm exp}~\Bigl\{g\int_0^{\beta}d\tau \int d^3{\bf x}({\tilde T}^a)_{\alpha\alpha}\delta^{(3)}({\bf x}-{\bf x}_0)
\Bigl(
(\del_3)^{-2}\bigl((D_k\Pi_A^k)^a +J_F^{a0}\bigr)
\Bigr)\Bigr\},
\label{eq:polyopdefaxial}
\end{equation}
where $({\tilde T}^a)_{\alpha\alpha}$ is defined in Eq.(\ref{eq:cartan}). It should be emphasized that the 
above operator is described by only the physical operators but not $A_{\tau}$. This is a main 
result of our paper. The above form of the operator might be guessed from Eqs.(\ref{eq:constraint}) and
(\ref{eq:dotakaxial}), but the proof that Eq.(\ref{eq:polyopdefaxial}) leads to the relation (\ref{eq:polyaxial2})
seems to be far from trivial, as we will see below.

For the operator (\ref{eq:polyopdefaxial}), the path integral representation for the left-hand-side of
Eq.(\ref{eq:polyaxial2}) is
\begin{eqnarray}
&&{\rm Tr}\left({\cal O}(\hat{\Pi}_A^{ak}, \hat{A}_k^a, \hat{\Pi}_{\psi}, \hat{\psi})
~\e^{-\beta H(\hat{\Pi}_A^k, \hat{A}_k, \hat{\Pi}_{\psi}, \hat{\psi})}\right)_{\rm axial}
=\sum_{\alpha=1}^D\int {\cal D}\Pi_A^{ak}{\cal D}A_k^a{\cal D}\Pi_{\psi}{\cal D}\psi 
\nonumber\\
&&\times ~{\rm exp}\int_0^{\beta}d\tau\int d^3{\bf x}
\Bigl\{g({\tilde T}^a)_{\alpha\alpha}\delta^{(3)}({\bf x}-{\bf x}_0)
\Bigl(
(\del_3)^{-2}\bigl((D_k\Pi_A^k)^a +J_F^{a0}\bigr)\Bigr)\nonumber\\
&&+i\Pi_A^{ak}\dot{A}_k^a + i\Pi_{\psi}\dot{\psi} -{\cal H}_{\rm axial}
\Bigr\},
\label{eq:polyexpaxial}
\end{eqnarray}
where the Hamiltonian ${\cal H}_{\rm axial}$ is given by Eq.(\ref{eq:hamiltonian}). 
Since the exponent of the right-hand-side in Eq.(\ref{eq:polyexpaxial}) is quadratic with 
respect $\Pi_A^{ak}$, whose explicit calculations and forms are given in the 
Appendix \ref{sec:polyaxialapp}, we can 
perform the Gaussian integration ${\cal D}\Pi_A^{ak}$, and we obtain that
\begin{eqnarray}
&&
{\rm Tr}\left({\cal O}(\hat{\Pi}_A^{ak}, \hat{A}_k^a, \hat{\Pi}_{\psi}, \hat{\psi})
~\e^{-\beta H(\hat{\Pi}_A^k, \hat{A}_k, \hat{\Pi}_{\psi}, \hat{\psi})}\right)_{\rm axial}
\nonumber\\
&&
=\sum_{\alpha=1}^D \int {\cal D}A_k^a {\cal D}\Pi_{\psi}{\cal D}\psi ~{\rm det}^{-\half}(M^{-1})~{\rm exp}
\int_0^{\beta}d\tau\int d^3{\bf x} \nonumber\\
&&\times ~\Bigl\{
-\half \dot{A}_k^a\dot{A}_l^a -\half (D_k\dot{A}_k)^a({\cal D}^{-1}D_l\dot{A}_l)^a
-i \dot{A}_k^a (D_k{\cal D}^{-1}J_F^0)^a +\half J_F^{a0}({\cal D}^{-1}J_F^0)^a
\nonumber
\\
&&
-\frac{1}{4}F_{kl}^aF_{kl}^a -\half \del_3A_k^a \del_3A_k^a
+i\Pi_{\psi}\dot{\psi} +{\bar\psi}i \gamma^kD_k\psi +{\bar\psi}i \gamma^3\del_3\psi
\nonumber\\
&&
+g({\tilde T}^a)_{\alpha\alpha}\delta^{(3)}({\bf x} -{\bf x}_0) ({\cal D}^{-1}J_F^0)^a 
+ig ({\tilde T}^a)_{\alpha\alpha}\delta^{(3)}({\bf x} -{\bf x}_0)({\cal D}^{-1}D_k\dot{A}_k)^a
\nonumber\\
&&-\frac{g^2}{2} \bigl(({\tilde T}^a)_{\alpha\alpha}\delta^{(3)}({\bf x} -{\bf x}_0)\bigr)
\bigl(
(\del_3)^{-2}({\tilde T}^a)_{\alpha\alpha}\delta^{(3)}({\bf x} -{\bf x}_0)
\bigr)
\nonumber\\
&&+\frac{g^2}{2} \bigl(({\tilde T}^a)_{\alpha\alpha}\delta^{(3)}({\bf x} -{\bf x}_0)\bigr)
\bigl(
({\cal D}^{-1})^{ab}({\tilde T}^b)_{\alpha\alpha}\delta^{(3)}({\bf x} -{\bf x}_0)
\bigr)
\Bigr\}.
\label{eq:polyexpaxial2}
\end{eqnarray}

We next insert the Gaussian integral, 
\begin{eqnarray}
&&1=\int {\cal D}A_{\tau}^a ~{\rm det}^{\half}({\cal D})~{\rm exp}\int_0^{\beta}d\tau \int d^3{\bf x}\nonumber\\
&&\Bigl\{
-\half \Bigl(
iA_{\tau}^a -({\cal D}^{-1})^{ac}\bigl(i (D_k\dot{A}_k)^c +J_F^{c0}  +
g ({\tilde T}^c)_{\alpha\alpha}\delta^{(3)}({\bf x}-{\bf x}_0) \bigr)\Bigr)
\nonumber\\
&&\times ({\cal D})^{ad}
\Bigl(iA_{\tau}^d -({\cal D}^{-1})^{de}\bigl(i (D_l\dot{A}_l)^e +J_F^{e0}  + 
g ({\tilde T}^e)_{\alpha\alpha}\delta^{(3)}({\bf x}-{\bf x}_0) \bigr)
\Bigr\}.
\label{eq:polyatauaxial}
\end{eqnarray}
into Eq.(\ref{eq:polyexpaxial2}) in order to recover the $A_{\tau}^a$ degree of freedom. 
Let us note that we add new terms of the point color charge densities $g({\tilde T}^c)_{\alpha\alpha}
\delta({\bf x}-{\bf x}_0)$, compared with Eq.(\ref{eq:gaussatauaxial}). If we expand the exponent of
Eq.(\ref{eq:polyatauaxial}), which is given by Eq.(\ref{eq:atauexpan}) in the Appendix \ref{sec:canonicalaxial}, we 
see that some of the terms in 
the exponent of Eq.(\ref{eq:polyexpaxial2}) are cancelled with those in Eq.(\ref{eq:polyatauaxial}). 
Then,  we have
\begin{eqnarray}
&&
{\rm Tr}\left({\cal O}(\hat{\Pi}_A^{ak}, \hat{A}_k^a, \hat{\Pi}_{\psi}, \hat{\psi})
~\e^{-\beta H(\hat{\Pi}_A^k, \hat{A}_k, \hat{\Pi}_{\psi}, \hat{\psi})}\right)_{\rm axial}
\nonumber\\
&&=\sum_{\alpha=1}^D \int {\cal D}A_{\tau}^a{\cal D}A_k^a {\cal D}\Pi_{\psi}{\cal D}\psi 
~{\rm det}^{-\half}(M^{-1}){\rm det}^{\half}({\cal D})~{\rm exp}
\int_0^{\beta}d\tau\int d^3{\bf x} \nonumber\\
&&\Bigl\{
-\half \dot{A}_k^a\dot{A}_k^a -\frac{1}{4}F_{kl}^aF_{kl}^a -\half \del_3A_k^a \del_3A_k^a
+i\Pi_{\psi}\dot{\psi} +{\bar\psi}i \gamma^kD_k\psi +{\bar\psi}i \gamma^3\del_3\psi
+ iA_{\tau}^aJ_F^{a0}
\nonumber\\
&&
-\half (D_kA_{\tau})^a(D_kA_{\tau})^a -\half \del_3A_{\tau}^a \del_3A_{\tau}^a 
+\dot{A}_k^a(D_kA_{\tau})^a \nonumber\\
&&+ ig A_{\tau}^a({\tilde T}^a)_{\alpha\alpha}\delta^{(3)}({\bf x} -{\bf x}_0)\nonumber\\
&&
-\frac{g^2}{2}\bigl(({\tilde T}^a)_{\alpha\alpha}\delta^{(3)}({\bf x} -{\bf x}_0)\bigr)
\bigl((\del_3)^{-2}({\tilde T}^a)_{\alpha\alpha}\delta^{(3)}({\bf x} -{\bf x}_0)
\bigr)
\Bigr\}.
\label{eq:axialpoly}
\end{eqnarray}
We note that $A_{\tau}^a$ is recovered as the auxiliary field through the Gaussian 
integral (\ref{eq:polyatauaxial}). The fourth line of Eq.(\ref{eq:axialpoly}) is just what we wanted 
and corresponds to the Polyakov loop in Eq.(\ref{eq:polyaxial2}). Since the term comes
from the cross terms in the Gaussian integral (\ref{eq:polyatauaxial}), it turns out to be
difficult to get the Polyakov loop in the original path-ordered form of Eq.(\ref{eq:polyaxial}) from the 
above procedure. This is the reason why we rewrite the Polyakov loop of Eq.(\ref{eq:polyaxial}) into
the abelian form given in Eq.(\ref{eq:polyaxial2}).

We finally restore the $A_3^a$ degree of freedom by Eq.(\ref{eq:a3delta}) and the 
terms in the exponent (\ref{eq:axialpoly}) is summarized into the covariant form of the Lagrangian. 
The determinants are evaluated as before like Eq.(\ref{eq:determinantaxial}). Then, we arrive at
\begin{eqnarray}
&&{\rm Tr}\left({\cal O}(\hat{\Pi}_A^{ak}, \hat{A}_k^a, \hat{\Pi}_{\psi}, \hat{\psi})
~\e^{-\beta H(\hat{\Pi}_A^k, \hat{A}_k, \hat{\Pi}_{\psi}, \hat{\psi})}\right)_{\rm axial}\nonumber\\
&=&\sum_{\alpha=1}^D 
\int {\cal D}A_{\mu}^a {\cal D}{\bar\psi}{\cal D}\psi~{\rm det}(\del_3)
\prod_{x, a}\delta(A_3^a)~
{\rm exp}~\Bigl\{ig \int_0^{\beta}d\tau A_\tau^a(\tau, {\bf x}_0)({\tilde T}^a)_{\alpha\alpha} \Bigr\}
\nonumber\\
&&
\times ~{\rm exp} \int_0^{\beta}d\tau\int d^3{\bf x}
\Bigl\{
\frac{-1}{4}F^a_{\mu\nu}F^a_{\mu\nu}  - {\bar\psi}i \gamma_{\mu}D_{\mu}\psi 
\Bigr\}.
\label{eq:polyaxialfinal}
\end{eqnarray}
Hence, we have proved Eq.(\ref{eq:polyaxial2}) for the operator $\hat{\cal O}$ defined 
by Eq.(\ref{eq:polyopdefaxial}). One can come back to the equation (\ref{eq:polyaxial}) by the 
inverse gauge transformations for Eqs.(\ref{eq:gtrf}) $\sim$ (\ref{eq:gtrf3}). One 
can say that we prove the equation (\ref{eq:aimshiki})  in the axial gauge
for the operator ${\cal O}$ defined by Eq.(\ref{eq:polyopdefaxial}).

Before closing this section, let us discuss the last term in Eq.(\ref{eq:axialpoly}), which is the divergent self-energy
of the point color charge densities due to introducing the Polyakov loop operator (\ref{eq:polyopdefaxial}). 
If we define a new operator $\hat{\cal O}_{\rm new}$ in such a way that 
we subtract the self-energy part from the beginning by the counter term, $\delta_{\rm self}^a$,
\begin{equation}
\hat{\cal O}_{\rm new} \equiv 
\sum_{\alpha=1}^D{\rm exp}~\Big\{g\int_0^{\beta}d\tau \int d^3{\bf x}({\tilde T}^a)_{\alpha\alpha}\delta^{(3)}({\bf x}-{\bf x}_0)
\Bigl(
(\del_3)^{-2}\bigl((D_k\Pi_A^k)^a +J_F^{a0}+\delta_{\rm self}^a\bigr) 
\Bigr)\Bigr\},
\label{eq:newoaxial}
\end{equation}
where 
\begin{equation}
\delta_{\rm self}^a\equiv\frac{g}{2}({\tilde T}^a)_{\alpha\alpha}\delta^{(3)}({\bf x} -{\bf x}_0),
\label{eq:countertermaxial}
\end{equation}
then, the divergent term does not appear in the expression (\ref{eq:axialpoly}). We note that the
term subtracted by the counter term, $(\partial_3)^{-2}\delta_{self}^a$, obviously does not depend 
on the field.
\section{Hamiltonian in Coulomb Gauge}
\label{sec:secfive}
In this section with the following two subsections we reformulate the analyses 
done in the previous sections, $2, 3$ and $4$ for the case of the Coulomb gauge.

The Coulomb gauge is given by 
\begin{equation}
\del_iA^{ai}=0\qquad (i=1,2,3).
\label{eq:coulomb}
\end{equation}
One of the components, say, $A_3^a$ is eliminated by Eq.(\ref{eq:coulomb}) like
\begin{equation}
A_3^a=-(\del_3)^{-1}\del_kA_k^a .
\label{eq:solvea3}
\end{equation}
We impose the Coulomb gauge on the $\nu=0$ component of the 
equation of motion (\ref{eq:eom}) and formally solve it
for $A_0^a$. Then, we have
\begin{equation}
A_0^a =({\cal D}^{-1})^{ab}\Bigl((C_k\dot{A}_k)^b + J_F^{b0}\Bigr)\qquad (k=1,2 ).
\label{eq:constcoul}
\end{equation}
The operator  $({\cal D}^{-1})^{ab}$ is defined as the inverse of $({\cal D})^{ab} \equiv (D_i^2)^{ab}$.
We find it useful to leave $A_3^a$ in the covariant derivative $(D_3)^{ab}$ as it is in order 
to perform calculations as clear as possible, keeping the 
equation (\ref{eq:solvea3}) in mind. $(C_k)^{ab}$ in Eq.(\ref{eq:constcoul}) is defined by
\begin{equation}
(C_k)^{ab}\equiv (D_k)^{ab} -(D_3)^{ab}(\del_3)^{-1}\del_k .
\label{eq:defc}
\end{equation}
The partial integration with respect to $(C_k)^{ab}$ yields a new 
operator, accompanying a minus sign,  given by
\begin{equation}
({\tilde C}_k)^{ab}\equiv (D_k)^{ab} -\del_k(\del_3)^{-1}(D_3)^{ab} 
\label{eq:deftildec}
\end{equation}
and vice versa.

Likewise the case for the axial gauge, we eliminate $A_0^a$ by Eq. (\ref{eq:constcoul}) and 
$A_3^a$ by Eq.(\ref{eq:solvea3}) in the 
Lagrangian (\ref{eq:lag}). Then, we obtain the 
Lagrangian ${\cal L}(A_k, \psi)_{\rm Coul}$ written in terms of only physical degrees of 
freedom, whose explicit form is presented by Eq.(\ref{eq:tildelagcoul}) in the 
Appendix \ref{sec:canonicalcoul}. Then, the canonical momentum for $A_k^a$ is given by
\begin{equation}
\Pi_A^{ak}\equiv\frac{\del {\cal L}(A_k, \psi)_{\rm Coul}}{\del \dot{A}_k^a}
=(N)_{kl}^{ab}\dot{A}_l^b - ({\tilde C}_k{\cal D}^{-1}J_F^0)^a,
\label{eq:momcoul}
\end{equation}
where 
\begin{equation}
(N)_{kl}^{ab}\equiv (\delta_{kl} +\del_k(\del_3)^{-2}\del_l)\delta^{ab} -({\tilde C}_k{\cal D}^{-1}C_l)^{ab}.
\label{eq:ndefcoul}
\end{equation}
We obtain from Eq.(\ref{eq:momcoul}) that
\begin{equation}
\dot{A}_k^a=(N^{-1})_{kl}^{ab}\Bigl(\Pi_A^{bl} + ({\tilde C}_l{\cal D}^{-1}J_F^0)^b\Bigr),
\label{eq:dotakcoul}
\end{equation}
where $(N^{-1})_{kl}^{ab}$ is defined as the inverse of $(N)_{kl}^{ab}$ and its explicit form is 
\begin{equation}
(N^{-1})_{kl}^{ab} = (\delta_{kl} -\del_k\Delta^{-1}\del_l)\delta^{ab} 
+\bigl({\tilde C}^{\prime}_k{\tilde \Delta}^{-1}C_l^{\prime}\bigr)^{ab},
\label{eq:ninverse}
\end{equation}
where 
\begin{eqnarray}
({\tilde C}^{\prime}_k)^{ab} 
&=&(D_k)^{ab} -\del_k\Delta^{-1}\del \cdot (D)^{ab}
\label{eq:ctildeprime}
\\
(C^{\prime}_k)^{ab} 
&=&(D_k)^{ab} -(D)^{ab}\cdot\del \Delta^{-1}\del_k,
\label{eq:cprime}
\\
{\tilde \Delta}^{-1} &\equiv &\bigl(D\cdot \del \Delta^{-1}\del\cdot D\bigr)^{-1} ,
\label{eq:tildedelta}
\\
\Delta^{-1} &\equiv& (\del_i^2)^{-1}.
\end{eqnarray}
We have also used the notation,
$\del \cdot D \equiv \del_i D_i$. Let us note that $\del\cdot D =D\cdot \del$ in the Coulomb gauge. 
We present the relations among the operators, (\ref{eq:defc}), (\ref{eq:deftildec}), (\ref{eq:ctildeprime}),
(\ref{eq:cprime}) and the proof of Eq.(\ref{eq:ninverse}) in the Appendix \ref{sec:canonicalcoul}.
As for the fermion, we take the right derivative with respect to the Grassmann variable, so that we have 
\begin{equation}
\Pi_{\psi}\equiv\frac{\del {{\cal L}}(A_k, \psi)_{\rm Coul}}{\del\dot{\psi}} ={\bar\psi}i\gamma^0.
\end{equation}

Then, the Hamiltonian is obtained as
\begin{eqnarray}
{\cal H}_{\rm Coul} &=& \Pi_A^{ak}\dot{A}_k^a +\Pi_{\psi}\dot{\psi} - {{\cal L}}(A_k, \psi)_{\rm Coul}
\nonumber\\
&=&
\half \Pi_A^{ak}(N^{-1}_{kl}\Pi_A^l)^a + \Pi_A^{ak}(N^{-1}_{kl}{\tilde C}_l{\cal D}^{-1}J_F^0)^a
\nonumber\\
&&+\half ({\tilde C}_k{\cal D}^{-1}J_F^0)(N^{-1}_{kl}{\tilde C}_l{\cal D}^{-1}J_F^0)^a 
-\half J_F^{a0}({\cal D}^{-1}J_F^0)^a
\nonumber\\
&&+\frac{1}{4}F_{kl}^a F_{kl}^a +\half F_{3k}^aF_{3k}^a - \bar\psi i \gamma^k D_k \psi 
-\bar\psi i \gamma^3D_3\psi .
\label{eq:hamiltoniancoul}
\end{eqnarray}
This Hamiltonian is written in terms of only physical degrees of freedom as it should. 
If we compare the Hamiltonian (\ref{eq:hamiltoniancoul}) with that in the axial 
gauge (\ref{eq:hamiltonian}), we observe the correspondences among 
the operators in the Hamiltonians such as $M^{-1} \leftrightarrow N^{-1}, D_k \leftrightarrow {\tilde C}_k$ even though their
explicit forms are quite different. The similarity between them may be the consequence of the fact that in 
the both cases the physical degrees of freedom are $A_k, \Pi_A^{ak}, \psi$ and $\Pi_{\psi}$ alone. 
The Hamiltonian (\ref{eq:hamiltoniancoul}) can be recast into another form, which is presented
by Eq.(\ref{eq:anohamcoul}) in the Appendix \ref{sec:canonicalcoul}, and the self-energy of the point color charge densities of
fermion in the Coulomb gauge becomes manifest. 
We shall use the Hamiltonian (\ref{eq:hamiltoniancoul}) in the discussions below. 
\subsection{Trace Formula in Coulomb Gauge at Finite Temperature}
In this section we would like to repeat the same analyses as the case of the axial gauge done in the section $3$.
We first show that the Hamiltonian (\ref{eq:hamiltoniancoul}) can reproduce the 
well-known path integral representation
for the partition function in the Coulomb gauge at finite temperature,
\begin{eqnarray}
 {\rm Tr}\left(\e^{-\beta H(\hat{\Pi}_A^k, \hat{A}_k, \hat{\Pi}_{\psi}, \hat{\psi})}\right)_{\rm Coul}
&=&\int {\cal D}A_{\mu}^a {\cal D}{\bar\psi}{\cal D}\psi~{\rm det}(\del_i D_i)
\prod_{x, a}\delta(\del_iA_i^a)
\nonumber\\
&&\times~
{\rm exp} \int_0^{\beta}d\tau\int d^3{\bf x}
\Bigl\{
\frac{-1}{4}F^a_{\mu\nu}F^a_{\mu\nu}  - {\bar\psi}i \gamma_{\mu}D_{\mu}\psi
\Bigr\}.
\label{eq:coulpart}
\end{eqnarray}
The discussions here may be too pedagogical again, but is helpful for the discussions in the next subsection.
The path integral representation for the left-hand-side of Eq.(\ref{eq:coulpart}) is
\begin{eqnarray}
&&{\rm Tr}\left(\e^{-\beta H(\hat{\Pi}_A^k, \hat{A}_k, \hat{\Pi}_{\psi}, \hat{\psi})}\right)_{\rm Coul}
\nonumber\\
&=&
\int {\cal D}\Pi_A^{ak}{\cal D}A_k^a{\cal D}\Pi_{\psi}{\cal D}\psi ~
{\rm exp}
\int_0^{\beta}d\tau \int d^3{\bf x}
\Bigl\{
i\Pi_{A}^{ak}\dot{A}_k^a + i\Pi_{\psi}\dot{\psi} -{\cal H}_{\rm Coul}
\Bigr\}.
\label{eq:coulexp}
\end{eqnarray}
Since the exponent in Eq.(\ref{eq:coulexp})  is the quadratic with respect to $\Pi_A^{ak}$, one can perform the 
Gaussian integration. The detailed expression for the quadratic form is given by Eq.(\ref{eq:quadracoul}) in the 
Appendix \ref{sec:canonicalcoul}. Then, we have
\begin{eqnarray}
&&
{\rm Tr}\left(\e^{-\beta H(\hat{\Pi}_A^k, \hat{A}_k, \hat{\Pi}_{\psi}, \hat{\psi})}\right)_{\rm Coul}
=\int {\cal D}A_k^a{\cal D}\Pi_{\psi}{\cal D}\psi ~{\rm det}^{-\half}(N^{-1})
~{\rm exp}\int_0^{\beta}d\tau\int d^3{\bf x} \nonumber\\
&&\times~\Bigl\{
-\half \dot{A}_k^a\dot{A}_k^a 
-\half  \bigl((\del_3)^{-1}\del_k\dot{A}_k^a\bigr)\bigl(  (\del_3)^{-1}\del_l\dot{A}_l^a\bigr)
+\half \dot{A}_k^a({\tilde C}_k{\cal D}^{-1}C_l\dot{A}_l)^a \nonumber\\
&&
-i\dot{A}_k^a({\tilde C}_k{\cal D}^{-1}J_F^0)^a 
+\half J_F^{a0}({\cal D}^{-1}J_F^0)^a 
\nonumber
\\
&&
-\frac{1}{4}F_{kl}^a F_{kl}^a -\half F_{3k}^aF_{3k}^a 
+i\Pi_{\psi}\dot{\psi}
+ \bar\psi i \gamma^k D_k \psi  +\bar\psi i \gamma^3D_3\psi 
\Bigr\}.
\label{eq:coulpiint}
\end{eqnarray}
We next consider the Gaussian integral given by
\begin{eqnarray}
1=\int {\cal D}A_{\tau}^a ~{\rm det}^{\half}({\cal D})~{\rm exp}\int_0^{\beta}d\tau \int d^3{\bf x}
\Bigl\{-\half 
\Bigl(iA_{\tau}^a -({\cal D}^{-1})^{ac}\bigl(i (C_k\dot{A}_k)^c +J_F^{c0}   \bigr)\Bigr)
\nonumber\\
\times~
({\cal D})^{ad}
\Bigl(iA_{\tau}^d -({\cal D}^{-1})^{de}\bigl(i (C_l\dot{A}_l)^e +J_F^{e0}   \bigr)\Bigr)\Bigr\}
\label{eq:gausscoul}
\end{eqnarray}
in order to restore the $A_{\tau}$ degree of freedom. Inserting the Gaussian 
integration (\ref{eq:gausscoul}) into Eq.(\ref{eq:coulpiint}), some of the terms in 
the exponents of Eqs.(\ref{eq:coulpiint}) and (\ref{eq:gausscoul}) are cancelled each other. 
We present the expansion of the exponent of Eq.(\ref{eq:gausscoul}) in 
Eq.(\ref{eq:ataucoul2}) in the Appendix \ref{sec:canonicalcoul}. Then, we have
 \begin{eqnarray}
&& {\rm Tr}\left(\e^{-\beta H(\hat{\Pi}_A^k, \hat{A}_k, \hat{\Pi}_{\psi}, \hat{\psi})}\right)_{\rm Coul}
 =\int {\cal D}A_{\tau}^a{\cal D}A_k^a{\cal D}\Pi_{\psi}{\cal D}\psi~{\rm det}^{-\half}(N^{-1})~
 {\rm det}^{\half}({\cal D})~\nonumber\\
 &&\times~{\rm exp}\int_0^{\beta}d\tau \int d^3{\bf x} 
\Bigl\{
-\half \dot{A}_k^a\dot{A}_k^a
-\half  \bigl((\del_3)^{-1}\del_k\dot{A}_k^a\bigr)\bigl(  (\del_3)^{-1}\del_l\dot{A}_l^a\bigr)
 - \frac{1}{4}F_{kl}^aF_{kl}^a -\half F_{3k}^a F_{3k}^a \nonumber\\
&&
-\half  (D_i A_{\tau})^a(D_i A_{\tau})^a 
+\dot{A}_k^a(D_kA_{\tau})^a -(D_3A_{\tau})^a((\del_3)^{-1}\del_k\dot{A}_k^a)
\nonumber\\
&&+i\Pi_{\psi}\dot{\psi}
+ \bar\psi i \gamma^k D_k \psi  +\bar\psi i \gamma^3D_3\psi + iA_{\tau}^a J_F^{a0} 
\Bigr\}.
\label{eq:coulatau}
\end{eqnarray}
As in the case of the axial gauge, $A_{\tau}$ is recovered as the auxiliary 
field through the Gaussian integral (\ref{eq:gausscoul}). In order to restore the 
$A_3^a$ degree of freedom, let us consider the identity,
\begin{equation}
1=\int {\cal D}A_3^a \prod_{x, a}\delta(A_3^a +(\del_3)^{-1}\del_kA_k^a)~
\Bigl(=\int {\cal D}A_3^a ~{\rm det}(\del_3)\prod_{x, a}\delta(\del_iA_i^a) 
\Bigr).
\label{eq:a3deltacoul}
\end{equation}
Then, the exponent (\ref{eq:coulatau}) is 
summarized into the covariant form of the Lagrangian thanks to the delta function in 
Eq.(\ref{eq:a3deltacoul}). And as shown in the Appendix \ref{sec:detcoul}, the determinants are evaluated as
\begin{equation}
{\rm det}^{-\half}(N^{-1}) {\rm det}^{\half}({\cal D}){\rm det}(\del_3) = {\rm det}(\del_iD_i).
\label{eq:determinantcoul}
\end{equation}
Hence, we obtain that
\begin{eqnarray}
{\rm Tr}\left(\e^{-\beta H(\hat{\Pi}_A^k, \hat{A}_k, \hat{\Pi}_{\psi}, \hat{\psi})}\right)_{\rm Coul}
&=&
\int {\cal D}A_{\mu}^a {\cal D}{\bar\psi}{\cal D}\psi~{\rm det}(\del_iD_i)
\prod_{x, a}\delta(\del_iA_i^a)\nonumber\\
&&\times~
{\rm exp} \int_0^{\beta}d\tau\int d^3{\bf x}
\Bigl\{
\frac{-1}{4}F^a_{\mu\nu}F^a_{\mu\nu}  - {\bar\psi}i \gamma_{\mu}D_{\mu}\psi
\Bigr\},
\label{eq:partcoulfinal}
\end{eqnarray}
where we have used ${\cal D}\Pi_{\psi}={\cal D}\bar\psi$ and defined $\gamma_{\tau}\equiv -i\gamma^0$.
We have finished to prove the equation (\ref{eq:coulpart}).
\subsection{Polyakov Loop in Coulomb Gauge}
Let us proceed to discuss the Polyakov loop in the Coulomb 
gauge \footnote{As a different approach from ours, a 
Polyakov loop in the temporal gauge has been discussed in Ref.\cite{rein} and treated as a Wilson line
of the Hosotani mechanism\cite{hosotani} by 
exchanging the role of a spacial coordinate and the Euclidean time.}. We no longer have the 
$A_{\tau}$ degree of freedom, from which the Polyakov loop is usually defined, in the operator
formalism of the Coulomb gauge. Nevertheless, it is natural to
expect that the operator written in terms of only physical degrees of freedom corresponding to 
the Polyakov loop should exist and to be defined.

Our aim is to present the operator $\hat{\cal O}$ satisfying 
the equation (\ref{eq:aimshiki}) for the case of the Coulomb gauge, that is, it satisfies 
\begin{eqnarray}
&&{\rm Tr}\left({{\cal O}}({\hat\Pi}_A^{ak}, {\hat A}_k^a,{\hat \Pi}_{\psi}, {\hat \psi})
~\e^{-\beta H(\hat{\Pi}_A^k, \hat{A}_k, \hat{\Pi}_{\psi}, \hat{\psi})}\right)_{\rm Coul}
\nonumber\\
&=&\sum_{\alpha=1}^D
\int {\cal D}A_{\mu}^a {\cal D}{\bar\psi}{\cal D}\psi~{\rm det}(\del_iD_i)
\prod_{x, a}\delta(\del_i A_i^a)~
{\rm exp}~\Bigl\{ig\int_0^{\beta}d\tau A_{\tau}^a(\tau,{\bf x}_0)({\tilde T}^a)_{\alpha\alpha}\Bigr\} \nonumber\\
&&
\times~{\rm exp} \int_0^{\beta}d\tau\int d^3{\bf x}
\Bigl\{
\frac{-1}{4}F^a_{\mu\nu}F^a_{\mu\nu}  - {\bar\psi}i \gamma_{\mu}D_{\mu}\psi
\Bigr\},
\label{eq:polycouldef}
\end{eqnarray}
where as in the case of the axial gauge the Polyakov loop is reduced to the $U(1)$ gauge sector 
of the $SU(N)$ gauge group by the gauge transformations (\ref{eq:gtrf}) $\sim$ (\ref{eq:gtrf3}). 
Let us note again that $\hat{\cal O}$ should be written in terms of only physical degrees of freedom.

We shall show that 
\begin{equation}
\hat{\cal O}\equiv 
\sum_{\alpha=1}^D{\rm exp}~\Bigl\{g\int_0^{\beta}
d\tau \int d^3{\bf x}({\tilde T}^a)_{\alpha\alpha}\delta^{(3)}({\bf x}-{\bf x}_0)
\Bigl(
({\tilde \Delta}^{-1})^{ab}\bigl((C_k^{\prime}\Pi_A^k)^b +J_F^{b0}+\delta_{\rm self}^b\bigr) 
\Bigr)\Bigr\}
\label{eq:coulodef}
\end{equation}
satisfies the equation (\ref{eq:polycouldef}). Here we have inserted the counter term defined by 
\begin{equation}
\delta_{\rm self}^b\equiv\frac{g}{2}({\tilde T}^b)_{\alpha\alpha}\delta^{(3)}({\bf x} -{\bf x}_0)
\label{eq:countertermcoul}
\end{equation}
in the definition (\ref{eq:coulodef}). Note that the counter term (\ref{eq:countertermcoul})
is the same form as 
the one for the axial gauge (\ref{eq:countertermaxial}) and it does not depend on fields.
$\delta^b_{\rm self}$ cancels the 
divergent self-energy of the point color charge densities arising from 
introducing the operator (\ref{eq:coulodef}), that is, the Polyakov loop operator as we will see below.
It may be worth while pointing out that the term subtracted by the counter term is
$(\tilde\Delta^{-1})^{ab}\delta_{self}^b$, so that, unlike the case for the axial 
gauge, it depends on the gauge field through the operator $\tilde\Delta^{-1}$.
The above form of the operator (\ref{eq:coulodef}) might be guessed from Eqs.(\ref{eq:constcoul}) and
(\ref{eq:dotakcoul}), but it seems to be far from trivial that the operator actually
satisfies the equation (\ref{eq:polycouldef}), as we will see below.

The path integral representation for the left-hand-side of Eq.(\ref{eq:polycouldef}) is
\begin{eqnarray}
&&{\rm Tr}\left({\cal O}({\hat\Pi}_A^{ak}, {\hat A}_k^a,{\hat \Pi}_{\psi}, {\hat \psi})~
\e^{-\beta H(\hat{\Pi}_A^k, \hat{A}_k, \hat{\Pi}_{\psi}, \hat{\psi})}\right)_{\rm Coul}
\nonumber\\
&=&\sum_{\alpha=1}^D\int {\cal D}\Pi_A^{ak}{\cal D}A_k^a{\cal D}\Pi_{\psi}{\cal D}\psi 
\nonumber\\
&&\times~{\rm exp}\int_0^{\beta}d\tau\int d^3{\bf x}
\Bigl\{g({\tilde T}^a)_{\alpha\alpha}\delta^{(3)}({\bf x}-{\bf x}_0)
\Bigl(
({\tilde \Delta}^{-1})^{ab}\bigl((C_k^{\prime}\Pi_A^k)^b +J_F^{b0}+\delta_{\rm self}^b\bigr)\Bigr)\nonumber\\
&&+i\Pi_A^{ak}\dot{A}_k^a + i\Pi_{\psi}\dot{\psi} -{\cal H}_{\rm Coul}
\Bigr\},
\label{eq:polyoexp}
\end{eqnarray}
where the Hamiltonian ${\cal H}_{\rm Coul}$ is given by Eq.(\ref{eq:hamiltoniancoul}). The exponent 
in Eq.(\ref{eq:polyoexp}) is quadratic 
with respect to $\Pi_A^{ak}$, so that we can perform the integration ${\cal D}\Pi_A^{ak}$.
The detailed calculations and the explicit quadratic form for the exponent in Eq.(\ref{eq:polyoexp}) are given 
in the Appendix \ref{sec:polycoul}. Then, we have
\begin{eqnarray}
&&{\rm Tr}\left({\cal O}({\hat\Pi}_A^{ak}, {\hat A}_k^a,{\hat \Pi}_{\psi}, {\hat \psi})~
\e^{-\beta H(\hat{\Pi}_A^k, \hat{A}_k, \hat{\Pi}_{\psi}, \hat{\psi})}\right)_{\rm Coul}
=\sum_{\alpha=1}^D \int {\cal D}A_k^a {\cal D}\Pi_{\psi}{\cal D}\psi ~{\rm det}^{-\half}(N^{-1})~
\nonumber\\
&&\times~{\rm exp}
\int_0^{\beta}d\tau\int d^3{\bf x} 
\Bigl\{
-\half \dot{A}_k^a\dot{A}_k^a 
-\half  \bigl((\del_3)^{-1}\del_k\dot{A}_k^a\bigr)\bigl(  (\del_3)^{-1}\del_l\dot{A}_l^a\bigr)
+\half \dot{A}_k^a({\tilde C}_k{\cal D}^{-1}C_l\dot{A}_l)^a \nonumber\\
&&
-i \dot{A}_k^a ({\tilde C}_k{\cal D}^{-1}J_F^0)^a +\half J_F^{a0}({\cal D}^{-1}J_F^0)^a
-\frac{1}{4}F_{kl}^aF_{kl}^a -\half F_{3k}^a F_{3k}^a\nonumber\\
&&+i\Pi_{\psi}\dot{\psi} +{\bar\psi}i \gamma^kD_k\psi +{\bar\psi}i \gamma^3D_3\psi
\nonumber\\
&&
+g({\tilde T}^a)_{\alpha\alpha}\delta^{(3)}({\bf x} -{\bf x}_0)
({\tilde \Delta}^{-1})^{ab}\delta_{\rm self}^b
\nonumber
\\
&&+g({\tilde T}^a)_{\alpha\alpha}\delta^{(3)}({\bf x} -{\bf x}_0) ({\cal D}^{-1}J_F^0)^a 
+ig({\tilde T}^a)_{\alpha\alpha}\delta^{(3)}({\bf x} -{\bf x}_0)({\cal D}^{-1}C_k\dot{A}_k)^a
\nonumber\\
&&
-\frac{g^2}{2}({\tilde T}^a)_{\alpha\alpha}\delta^{(3)}({\bf x} -{\bf x}_0)
\bigl(({\tilde \Delta}^{-1})^{ab}({\tilde T}^b)_{\alpha\alpha}\delta^{(3)}({\bf x} -{\bf x}_0)\bigr)
\nonumber\\
&&+\frac{g^2}{2}({\tilde T}^a)_{\alpha\alpha}\delta^{(3)}({\bf x} -{\bf x}_0)
\bigl(({\cal D}^{-1})^{ab}({\tilde T}^b)_{\alpha\alpha}\delta^{(3)}({\bf x} -{\bf x}_0)\bigr)
\Bigr\}.
\label{eq:polyexpcoul2}
\end{eqnarray}

In order to recover the $A_{\tau}^a$ degree of freedom, we insert the Gaussian 
integral defined by 
\begin{eqnarray}
&&1=\int {\cal D}A_{\tau}^a ~{\rm det}^{\half}({\cal D})~{\rm exp}\int_0^{\beta}d\tau \int d^3{\bf x}\nonumber\\
&&\times~\Bigl\{-\half 
\Bigl(iA_{\tau}^a -({\cal D}^{-1})^{ac}\bigl(i (C_k\dot{A}_k)^c +J_F^{c0}  +
g ({\tilde T}^c)_{\alpha\alpha}\delta^{(3)}({\bf x}-{\bf x}_0) \bigr)\Bigr)
\nonumber\\
&&~~\times ({\cal D})^{ad}
\Bigl(iA_{\tau}^d -({\cal D}^{-1})^{de}\bigl(i (C_l\dot{A}_l)^e +J_F^{e0}  + 
g ({\tilde T}^e)_{\alpha\alpha}\delta^{(3)}({\bf x}-{\bf x}_0) \bigr)\Bigr)\Bigr\}.
\label{eq:polyataucoul}
\end{eqnarray}
into Eq.(\ref{eq:polyexpcoul2}). We have added new terms of the point color charge 
densities $g({\tilde T}^c)_{\alpha\alpha} \delta({\bf x}-{\bf x}_0)$, compared with Eq.(\ref{eq:gausscoul}).
In Eq.(\ref{eq:polyataucoul2}) of the Appendix \ref{sec:canonicalcoul}, we explicitly 
present the  expansion of the exponent of Eq.(\ref{eq:polyataucoul}) and 
we find that some of the terms in the exponent of Eq.(\ref{eq:polyataucoul}) are cancelled with 
those in Eq.(\ref{eq:polyexpcoul2}). Then,  we obtain that
\begin{eqnarray}
&&{\rm Tr}\left({\cal O}({\hat\Pi}_A^{ak}, {\hat A}_k^a,{\hat \Pi}_{\psi}, {\hat \psi})~
\e^{-\beta H(\hat{\Pi}_A^k, \hat{A}_k, \hat{\Pi}_{\psi}, \hat{\psi})}\right)_{\rm Coul}
\nonumber\\
&&=\sum_{\alpha=1}^D \int {\cal D}A_{\tau}^a {\cal D}A_k^a
 {\cal D}\Pi_{\psi}{\cal D}\psi ~{\rm det}^{-\half}(N^{-1}){\rm det}^{\half}({\cal D})~
{\rm exp}\int_0^{\beta}d\tau\int d^3{\bf x} 
\nonumber\\
&&\times~
\Bigl\{
-\half \dot{A}_k^a\dot{A}_k^a 
-\half  \bigl((\del_3)^{-1}\del_k\dot{A}_k^a\bigr)\bigl(  (\del_3)^{-1}\del_l\dot{A}_l^a\bigr)
-\frac{1}{4}F_{kl}^aF_{kl}^a -\half F_{3k}^a F_{3k}^a
\nonumber\\
&&
-\half (D_iA_{\tau})^a(D_iA_{\tau})^a 
+\dot{A}_k^a(D_kA_{\tau})^a - (D_3A_{\tau})^a((\del_3)^{-1}\del_k\dot{A}_k^a)
\nonumber\\
&&
+i\Pi_{\psi}\dot{\psi} +{\bar\psi}i \gamma^kD_k\psi +{\bar\psi}i \gamma^3D_3\psi
+ iA_{\tau}^aJ_F^{a0}
+ ig A_{\tau}^a({\tilde T}^a)_{\alpha\alpha}\delta^{(3)}({\bf x} -{\bf x}_0)
\nonumber\\
&&+g({\tilde T}^a)_{\alpha\alpha}\delta^{(3)}({\bf x} -{\bf x}_0)
({\tilde \Delta}^{-1})^{ab}\delta_{\rm self}^b\nonumber\\
&&
-\frac{g^2}{2}({\tilde T}^a)_{\alpha\alpha}\delta^{(3)}({\bf x} -{\bf x}_0)
\bigl(({\tilde \Delta}^{-1})^{ab}({\tilde T}^b)_{\alpha\alpha}\delta^{(3)}({\bf x} -{\bf x}_0)\bigr)
\label{eq:coulpoly}
\Bigr\}.
\end{eqnarray}
The last two terms are found to be canceled with the choice of Eq.(\ref{eq:countertermcoul}). 
The $A_{\tau}$ degree of freedom is recovered as the auxiliary field through the Gaussian 
integral (\ref{eq:polyataucoul}). The last term in the fourth line of Eq.(\ref{eq:coulpoly}) just corresponds to 
the Polyakov loop in Eq.(\ref{eq:polycouldef}).

We finally restore the $A_3^a$ degree of freedom by Eq.(\ref{eq:a3deltacoul}). 
As before, the exponent is summarized into the covariant 
form of the Lagrangian and the determinant is evaluated as Eq.(\ref{eq:determinantcoul}).
The result is 
\begin{eqnarray}
&&{\rm Tr}\left({\cal O}({\hat\Pi}_A^{ak}, {\hat A}_k^a,{\hat \Pi}_{\psi}, {\hat \psi})~
\e^{-\beta H(\hat{\Pi}_A^k, \hat{A}_k, \hat{\Pi}_{\psi}, \hat{\psi})}\right)_{\rm Coul}
=\sum_{\alpha=1}^D 
\int {\cal D}A_{\mu}^a {\cal D}{\bar\psi}{\cal D}\psi~{\rm det}(\del_iD_i)
\prod_{x, a}\delta(\del_i A_i^a)\nonumber\\
&&\times ~{\rm exp}\Bigl\{ig \int_0^{\beta}d\tau A_\tau^{\tilde a}(\tau, {\bf x}_0)({\tilde T}^a)_{\alpha\alpha} \Bigr\}
{\rm exp} \int_0^{\beta}d\tau\int d^3{\bf x}
\Bigl\{
\frac{-1}{4}F^a_{\mu\nu}F^a_{\mu\nu}  - {\bar\psi}i \gamma_{\mu}D_{\mu}\psi 
\Bigr\}.
\label{eq:polycoulfinal}
\end{eqnarray}
Thus, we have proved the equation (\ref{eq:polycouldef}) and 
it implies that we have verified the equation (\ref{eq:aimshiki}) in the Coulomb gauge
for the operator $\hat{\cal O}$ defined by Eq.(\ref{eq:coulodef}). 

It may be worth here mentioning the equivalence between the axial gauge
and the Coulomb gauge. The Faddeev-Popov method guarantees the
equivalence among different gauge choices in the path integral formalism
even at finite temperature. 
Although the equivalence is believed to be true even in the operator
formalism at finite temperature, the operator correspondence between
operator formalisms with different gauge choices seems to be less obvious.
Actually, our final results (\ref{eq:polyaxialfinal}) and (\ref{eq:polycoulfinal}) show
that the right-hand-sides of them are mutually transformed by the
Faddeev-Popov manner. This fact implies that the results may prove
indirectly the equivalence between the axial gauge and the Coulomb gauge
in the operator formalism at finite temperature
with the corresponding Polyakov loop operators.
\section{Conclusions}
In quantizing the non-abelian gauge theory with the non-covariant gauge fixing such as the 
axial and Coulomb gauge, any physical operator such as the Hamiltonian is written in
terms of only physical degrees of freedom, by which the Hilbert space is spanned. 
$A_{0}$ is not dynamical variable, so that it disappears in the Hilbert space. Then, one may wonder 
how one should define the operator, written in terms of only physical
degrees of freedom, corresponding to the Polyakov loop operator 
when one considers the theory in these gauge fixings.

In order to answer the question, we have first studied the canonical formulation of the 
non-abelian gauge theory in the axial and Coulomb gauge, in which the physical degrees of freedom are
clarified. Namely, we emphasis on starting with the 
Lagrangian obtained by imposing the gauge condition and by eliminating $A_0^a$ through the 
equation of motion, from which the canonical momentum for the gauge field is defined.
The Hamiltonian obtained in this way is written in terms of only
physical degrees of freedom as it should though it has the complicated form including the non-local
operator.

We have constructed the operator corresponding to 
the Polyakov loop operator in terms of only physical degrees 
of freedom in the axial and Coulomb gauge. In order to confirm that the defined operator 
is actually the standard Polyakov loop operator,  we have evaluated the trace formula 
for the operator at finite temperature and showed that it can be rewritten into the covariant path integral form 
for the usual Polyakov loop operator with all the gauge degrees of freedom under the chosen gauge fixing.

We have encountered the divergent quantity in the process 
to obtain the covariant path integral form. It is the self-energy of the point 
color charge densities owing to introducing the operator $\hat{\cal O}$, which corresponds to the Polyakov loop.
Introducing the Polyakov loop is essentially the same with considering 
the point color charge densities of fermion. This manifestly appears in the 
Hamiltonians, (\ref{eq:anoham}) and (\ref{eq:anohamcoul}) in the Appendix. In the case of the axial gauge, the
divergent self-energy does not depend on the field, while in the case of the Coulomb gauge it does.
One can subtract the self-energy by introducing the counter term, $\delta_{\rm self}^a$
in the definition of the operator $\hat{\cal O}$, and the self-energy does not appear in the final
expression of the covariant path integral form.

%
%

In proposing the operators (\ref{eq:newoaxial}) and (\ref{eq:coulodef}), we have taken into 
account of the fact that the Polyakov loop operator has the same physical effect of
introducing a color charge density. In fact, we have guessed the forms of 
the operators (\ref{eq:newoaxial}) and  (\ref{eq:coulodef}) from the second and third terms in the 
Hamiltonians  (\ref{eq:anoham}) and  (\ref{eq:anohamcoul}).
%
%
It is also very important and interesting to investigate the theoretical aspects 
of the non-abelian gauge theory at finite temperature based on the present 
work in addition to give the complete physical interpretation of the operator.
%
\begin{center}
{\bf Acknowledgement}
\end{center}
The authors thank Norisuke Sakai for raising the issue discussed in this paper.
This work is supported in part by Grants-in-Aid for Scientific 
Research [No.~15K05055 and No.~25400260 (M.S.)] from the Ministry of 
Education, Culture, Sports, Science and Technology (MEXT) in Japan.
%
%
\appendix
\section{Axial gauge}
\label{sec:canonicalaxial}
In this Appendix we present some details of calculations and necessary formulae in order to
derive the results in the text.
\subsection{Hamiltonian in axial gauge}
\label{sec:canohamaxial}
The Lagrangian ${\cal L}(A_k^a, \psi)_{\rm axial}$ is obtained by imposing the axial gauge $A_3^a=0$ 
and eliminating $A_0^a$ by using the constraint (\ref{eq:constraint}) in the Lagrangian (\ref{eq:lag}). 
Straightforward calculations yield 
\begin{eqnarray}
{\cal L}(A_k^a, \psi)_{\rm axial}&=&
\half \dot{A}_k^a (M_{kl}\dot{A}_l)^a
-\dot{A}_k^a(D_k{\cal D}^{-1} J_F^0)^a
+\half J_F^{a0}({\cal D}^{-1}J_F^0)^a
\nonumber\\
&&-\frac{1}{4}F_{kl}^aF_{kl}^a -\half \del_3A_k^a \del_3 A_k^a
+{\bar\psi}i\gamma^0\dot{\psi} +{\bar\psi}i\gamma^kD_k\psi +{\bar\psi}i\gamma^3\del_3\psi ,
\label{eq:tildelagaxial}
\end{eqnarray}
where we have used the definition for $(M)_{kl}^{ab}$, (\ref{eq:mexplicit}). In deriving (\ref{eq:tildelagaxial}), 
we have formally performed the partial integration with respect to $D_k$ and $\cal D$.
The partial integration with respect to $D_k$ accompanies a minus sign, while that with
respect to $\cal D$ does not. And we also note that the partial integration with respect to 
$(M)_{kl}^{ab}, (M^{-1})_{kl}^{ab}$ does not accompany a minus sign.

As mentioned in the section $2$, the Hamiltonian (\ref{eq:hamiltonian}),
\begin{eqnarray}
{\cal H}_{\rm axial} &=& \Pi_A^{ak}\dot{A}_k^a +\Pi_{\psi}\dot{\psi} - {{\cal L}}(A_k, \psi)_{\rm axial} 
\nonumber\\
&=&
\half \Pi_A^{ak}(M^{-1}_{kl}\Pi_A^l)^a + \Pi_A^{ak}(M^{-1}_{kl}D_l{\cal D}^{-1}J_F^0)^a
\nonumber\\
&&+\half (D_k{\cal D}^{-1}J_F^0)^a(M^{-1}_{kl}D_l{\cal D}^{-1}J_F^0)^a -\half J_F^{a0}({\cal D}^{-1}J_F^0)^a
\nonumber\\
&&+\frac{1}{4}F_{kl}^a F_{kl}^a +\half \del_3A_k^a\del_3A_k^a - \bar\psi i \gamma^k D_k \psi 
-\bar\psi i \gamma^3\del_3\psi 
\label{eq:hamiltonianapp}
\end{eqnarray}
can be rewritten into another form, which is obtained by 
using the explicit form of $(M^{-1})_{kl}^{ab}$ for the second and the third 
term in Eq.(\ref{eq:hamiltonianapp}) after the partial integration 
with respect to $M_{kl}^{-1}, D_k$ and ${\cal D}^{-1}$. These two terms are written as
\begin{eqnarray}
&&-({\cal D}^{-1}D_lM^{-1}_{lk}\Pi_A^{k})^aJ_F^{a0}
-\half J_F^{a0}({\cal D}^{-1}D_kM^{-1}_{kl}D_l{\cal D}^{-1}J_F^0)^a 
\nonumber\\
&&=\Pi_A^{ak}\Bigl(D_k(\del_3)^{-2}J_F^0\Bigr)^a -
\half J_F^{a0}\Big(\big((\del_3)^{-2}\delta^{ab} - ({\cal D}^{-1})^{ab}\bigr)J_F^{b0}\Bigr) ,
\label{eq:relaxialapp}
\end{eqnarray}
where we have used the relations,
\begin{eqnarray}
({\cal D}^{-1}D_lM^{-1}_{lk})^{ab}=(\del_3)^{-2} (D_k)^{ab}
%
%
\label{eq:newrelaxial1}
\end{eqnarray}
and 
\begin{eqnarray}
\Bigl({\cal D}^{-1}D_kM_{kl}^{-1}D_l{\cal D}^{-1}\Bigr)^{ab}
=(\del_3)^{-2}\delta^{ab} -({\cal D}^{-1})^{ab},
%
%
%
\label{eq:newrelaxial2}
\end{eqnarray}
which can be easily shown by using the explicit form of $(M^{-1})_{kl}^{ab}$ given by
Eq.(\ref{eq:minv}). We have finally performed the partial integration
with respect to $D_k$ and $(\del_3)^{-2}$ in order to  have the result (\ref{eq:relaxialapp}).
Then, we obtain that
\begin{eqnarray}
{\cal H}_{\rm axial} &=&
\half \Pi_A^{ak}(M^{-1}_{kl}\Pi_A^l)^a + \Pi_A^{ak}(D_k(\del_3)^{-2}J_F^0)^a
-\half J_F^{a0}(\del_3)^{-2}J_F^{a0}\nonumber\\
&&+\frac{1}{4}F_{kl}^a F_{kl}^a +\half \del_3A_k^a\del_3A_k^a - \bar\psi i \gamma^k D_k \psi 
-\bar\psi i \gamma^3\del_3\psi .
\label{eq:anoham}
\end{eqnarray}
Let us note that the third term in the first line in Eq.(\ref{eq:anoham}) is well-known to be the
self-energy of the point color charge densities of fermion in the axial gauge.
The self-energy manifestly appears in this form of the Hamiltonian.
\subsection{Trace formula in axial gauge}
\label{sec:traceaxial}
In order to perform the integration ${\cal D}\Pi_A^{ak}$, we complete the square 
with respect to $\Pi_A^{ak}$ in the exponent of Eq. (\ref{eq:partaxial}),
\begin{eqnarray}
&&i\Pi_A^{ak}\dot{A}_k^a +i \Pi_{\psi}\dot{\psi} -{\cal H}_{\rm axial}\nonumber\\
&=&-\half 
\Bigl(\Pi_A^{ak} -i (M_{kl}\dot{A}_l)^a +(D_k {\cal D}^{-1}J_F^0)^a\Bigr)
(M^{-1})_{km}^{ab}\Bigl(\Pi_A^{bm} -i (M_{mn}\dot{A}_n)^b +(D_m {\cal D}^{-1}J_F^0)^b\Bigr)
\nonumber\\
&&
-\half \dot{A}_k^a(M_{kl}\dot{A}_l)^a -i\dot{A}_k^a(D_k{\cal D}^{-1}J_F^0)^a 
+\half J_F^{a0}({\cal D}^{-1}J_F^0)^a 
-\frac{1}{4}F_{kl}^a F_{kl}^a -\half \del_3A_k^a\del_3A_k^a \nonumber\\
&&
+i\Pi_{\psi}\dot{\psi}+ \bar\psi i \gamma^k D_k \psi  +\bar\psi i \gamma^3\del_3\psi .
\label{eq:quadraaxial}
\end{eqnarray}
The first term in the third line of Eq.(\ref{eq:quadraaxial}) can be written as
\begin{equation}
-\half \dot{A}_k^a(M_{kl}\dot{A}_l)^a=-\half \dot{A}_k^a\dot{A}_k^a -\half (D_k\dot{A}_k)^a({\cal D}^{-1}D_l\dot{A}_l)^a 
\label{eq:adot2axial}
\end{equation}
by using the explicit form for $(M)_{kl}^{ab}$, (\ref{eq:mexplicit}) and by performing the partial integration 
for $D_k$. Then, we obtain the equation (\ref{eq:afterdpi}) in the text.
If we expand the exponent of the Gaussian integral (\ref{eq:gaussatauaxial}), we have, after
performing the partial integration with respect to ${\cal D}$, 
\begin{eqnarray}
&&-\half 
\Bigl(iA_{\tau}^a -({\cal D}^{-1})^{ac}\bigl(i (D_k\dot{A}_k)^c +J_F^{c0}   \bigr)\Bigr)
({\cal D})^{ad}
\Bigl(iA_{\tau}^d -({\cal D}^{-1})^{de}\bigl(i (D_l\dot{A}_l)^e +J_F^{e0}   \bigr)\Bigr)
\nonumber\\
&=&-\half  (D_k A_{\tau})^a(D_k A_{\tau})^a 
-\half \del_3A_{\tau}^a \del_3A_{\tau}^a + iA_{\tau}^a J_F^{a0} + \dot{A}_k^a(D_kA_{\tau})^a
\nonumber\\
&&+\half (D_k\dot{A}_k)^a({\cal D}^{-1}D_l\dot{A}_l)^a
+i \dot{A}_k^a(D_k{\cal D}^{-1}J_F^0)^a -\half J_F^{a0}({\cal D}^{-1}J_F^0)^a .
\label{eq:expansionaxial}
\end{eqnarray}
We observe that some of the terms in the second line of the right-hand-side of Eq. (\ref{eq:quadraaxial}) 
and the right-hand-side of Eq.(\ref{eq:expansionaxial}) are canceled each other to yield the 
result (\ref{eq:axialprefinal}) in the text.
\subsection{Polyakov loop in axial gauge}
\label{sec:polyaxialapp}
In order to perform the integration ${\cal D}\Pi_A^{ak}$, we need to complete the square with respect to $\Pi_A^{ak}$
in the exponent of Eq.(\ref{eq:polyexpaxial}), which are given by
\begin{eqnarray}
&&-\half 
\Bigl(\Pi_A^{ak} - (M)_{kl}^{ab}\bigl(i\dot{A}_l^b  - g(D_l)^{bd}(\del_3)^{-2}({\tilde T}^d)_{\alpha\alpha}
\delta^{(3)}({\bf x} -{\bf x}_0)\bigr)
+(D_k {\cal D}^{-1}J_F^0)^b\Bigr)(M^{-1})_{km}^{ag}
\nonumber\\
&&\times 
\Bigl(\Pi_A^{gm} - (M)_{mn}^{gc}\bigl(i\dot{A}_n^c 
- g(D_n)^{cf}(\del_3)^{-2}({\tilde T}^f)_{\alpha\alpha}\delta^{(3)}({\bf x} -{\bf x}_0)\bigr)
+(D_m {\cal D}^{-1}J_F^0)^g\Bigr).
\end{eqnarray}
There are new terms containing the gauge coupling $g$ due to introducing the 
operator (\ref{eq:polyopdefaxial}), which do not appear in Eq.(\ref{eq:quadraaxial}).
According to the perfect square with respect to $\Pi_A^{ak}$, we should add the following terms,
\begin{eqnarray}
&&-g({\tilde T}^a)_{\alpha\alpha} \delta^{(3)}({\bf x}-{\bf x}_0)\bigl((\del_3)^{-2}(D_k\Pi_A^k)^a\bigr)
+ig({\tilde T}^a)_{\alpha\alpha} \delta^{(3)}({\bf x}-{\bf x}_0)\bigl((\del_3)^{-2}D_kM_{kl}\dot{A}_l\bigr)^a
\nonumber\\
&&+g(M)_{kl}^{ab}(D_l)^{bd}(\del_3)^{-2}({\tilde T}^d)_{\alpha\alpha}\delta^{(3)}({\bf x}-{\bf x}_0)
\bigl((M^{-1})_{km}^{ag}(D_m{\cal D}^{-1}J_F^0)^g\bigr)
\nonumber\\
&&
-\frac{g^2}{2} \bigl(({\tilde T}^a)_{\alpha\alpha}\delta^{(3)}({\bf x} -{\bf x}_0)\bigr)
\bigl((\del_3)^{-2}
(D_kM_{kl}D_l)^{ab}(\del_3)^{-2}({\tilde T}^b)_{\alpha\alpha}\delta^{(3)}({\bf x} -{\bf x}_0)
\bigr),
\label{eq:axialnewrel0}
\end{eqnarray}
where we have performed the partial integration with respect 
to $(\del_3)^{-2}, D_k$ and $(M)_{kl}^{ab}$. They can be rewritten into the simple forms by using 
the explicit form of $M_{kl}^{ab}$ given by Eq.(\ref{eq:mexplicit}). It is easy to check that 
\begin{eqnarray}
(D_lM_{lk})^{ab}&=&(D_l)^{ac}\bigl(\delta^{cb}\delta_{lk}-(D_l{\cal D}^{-1}D_k)^{cb}\bigr)
=(D_k)^{ab} - (D_l^2{\cal D}^{-1}D_k)^{ab}\nonumber\\
&=&(D_k)^{ab} - \bigl(({\cal D}-(\del_3)^2){\cal D}^{-1}D_k\bigr)^{ab}
=(\del_3)^2({\cal D}^{-1}D_k)^{ab}
\label{eq:axialnewrel1}
\end{eqnarray}
and 
\begin{eqnarray}
(D_kM_{kl}D_l)^{ab}&=&
(D_kM_{kl})^{ac}(D_l)^{cb}\stackrel{\rm (\ref{eq:axialnewrel1})}{=}(\del_3)^2({\cal D}^{-1}D_l)^{ac}(D_l)^{cb}
\nonumber\\
&=&(\del_3)^2\bigl({\cal D}^{-1}({\cal D}-(\del_3)^2)\bigr)^{ab}
=(D_k^2)^{ab}-(D_k^2{\cal D}^{-1}D_l^2)^{ab}\nonumber\\
&=&(\del_3)^2\delta^{ab} - (\del_3)^2({\cal D}^{-1})^{ab}(\del_3)^2 .
\label{eq:axialnewrel2}
\end{eqnarray}
Then, the equation (\ref{eq:axialnewrel0}) becomes
\begin{eqnarray}
&&-g({\tilde T}^a)_{\alpha\alpha} \delta^{(3)}({\bf x}-{\bf x}_0)\bigl((\del_3)^{-2}(D_k\Pi_A^k)^a\bigr)
+ig({\tilde T}^a)_{\alpha\alpha} \delta^{(3)}({\bf x}-{\bf x}_0)\bigl((\del_3)^{-2}D_kM_{kl}\dot{A}_l\bigr)^a\nonumber\\
&&+g(M)_{kl}^{ab}(D_l)^{bd}(\del_3)^{-2}({\tilde T}^d)_{\alpha\alpha}\delta^{(3)}({\bf x}-{\bf x}_0)
\bigl((M^{-1})_{km}^{ag}(D_m{\cal D}^{-1}J_F^0)^g\bigr)
\nonumber\\
&&
-\frac{g^2}{2} \bigl(({\tilde T}^a)_{\alpha\alpha}\delta^{(3)}({\bf x} -{\bf x}_0)\bigr)
\bigl((\del_3)^{-2}
(D_kM_{kl}D_l)^{ab}(\del_3)^{-2}({\tilde T}^b)_{\alpha\alpha}\delta^{(3)}({\bf x} -{\bf x}_0)\bigr)
\nonumber\\
&=&
-g\bigl(({\tilde T}^a)_{\alpha\alpha}\delta^{(3)}({\bf x}-{\bf x}_0)\bigr)
\Bigl((\del_3)^{-2}\bigl((D_k\Pi_A^k)^a +J_F^0\bigr)^a\Bigr)
\nonumber\\
&&
+g\bigl(({\tilde T}^a)_{\alpha\alpha}\delta^{(3)}({\bf x}-{\bf x}_0)\bigr)\bigl({\cal D}^{-1}J_F^0\bigr)^a
+
ig({\tilde T}^a)_{\alpha\alpha} \delta^{(3)}({\bf x}-{\bf x}_0)({\cal D}^{-1}D_k\dot{A}_k)^{a}
\nonumber\\
&&
-\frac{g^2}{2} \bigl(({\tilde T}^a)_{\alpha\alpha}\delta^{(3)}({\bf x} -{\bf x}_0)\bigr)
\bigl((\del_3)^{-2}({\tilde T}^b)_{\alpha\alpha}\delta^{(3)}({\bf x} -{\bf x}_0)\bigr)
\nonumber\\
&&+\frac{g^2}{2} \bigl(({\tilde T}^a)_{\alpha\alpha}\delta^{(3)}({\bf x} -{\bf x}_0)\bigr)
\bigl(({\cal D}^{-1})^{ab}({\tilde T}^b)_{\alpha\alpha}\delta^{(3)}({\bf x} -{\bf x}_0)\bigr),
\label{eq:axialnewrel3}
\end{eqnarray}
where the first term, aside from the minus sign, in the right-hand-side is nothing but the exponent of 
the operator (\ref{eq:polyopdefaxial}). Therefore, the 
exponent of the right-hand-side of Eq.(\ref{eq:polyexpaxial}) is written into the quadratic form 
with respect to $\Pi_A^{ak}$ as
\begin{eqnarray}
&&
g({\tilde T}^a)_{\alpha\alpha}\delta^{(3)}({\bf x}-{\bf x}_0)
\Bigl(
(\del_3)^{-2}\bigl((D_k\Pi_A^k)^a +J_F^{a0}\bigr)
\Bigr) +
i\Pi_A^{ak}\dot{A}_k^a +i \Pi_{\psi}\dot{\psi} -{\cal H}_{\rm axial}
\nonumber\\
&=&
-\half\Bigl(\Pi_A^{ak} - (M)_{kl}^{ab}\bigl(i\dot{A}_l^b  - g(D_l)^{bd}(\del_3)^{-2}({\tilde T}^d)_{\alpha\alpha}
\delta^{(3)}({\bf x} -{\bf x}_0)\bigr)+(D_k {\cal D}^{-1}J_F^0)^b\Bigr)(M^{-1})_{km}^{ag}
\nonumber\\
&&\times 
\Bigl(\Pi_A^{gm} - (M)_{mn}^{gc}\bigl(i\dot{A}_n^c 
- g(D_n)^{cf}(\del_3)^{-2}({\tilde T}^f)_{\alpha\alpha}\delta^{(3)}({\bf x} -{\bf x}_0)\bigr)
+(D_m {\cal D}^{-1}J_F^0)^g\Bigr)
\nonumber\\
&&
-\half \dot{A}_k^a(M_{kl}\dot{A}_l)^a -i\dot{A}_k^a(D_k{\cal D}^{-1}J_F^0)^a 
+\half J_F^{a0}({\cal D}^{-1}J_F^0)^a 
-\frac{1}{4}F_{kl}^a F_{kl}^a -\half \del_3A_k^a\del_3A_k^a 
\nonumber\\
&&+i\Pi_{\psi}\dot{\psi}+ \bar\psi i \gamma^k D_k \psi  +\bar\psi i \gamma^3\del_3\psi 
\nonumber\\
&&
+g({\tilde T}^a)_{\alpha\alpha}\delta^{(3)}({\bf x} -{\bf x}_0)({\cal D}^{-1}J_F^0)^a 
+ig ({\tilde T}^a)_{\alpha\alpha}\delta^{(3)}({\bf x} -{\bf x}_0)({\cal D}^{-1}D_k\dot{A}_k)^a
\nonumber\\
&&
-\frac{g^2}{2} \bigl(({\tilde T}^a)_{\alpha\alpha}\delta^{(3)}({\bf x} -{\bf x}_0)\bigr)
\bigl((\del_3)^{-2}({\tilde T}^a)_{\alpha\alpha}\delta^{(3)}({\bf x} -{\bf x}_0)\bigr)
\nonumber\\
&&
+\frac{g^2}{2} \bigl(({\tilde T}^a)_{\alpha\alpha}\delta^{(3)}({\bf x} -{\bf x}_0)\bigr)
\bigl(({\cal D}^{-1})^{ab}({\tilde T}^b)_{\alpha\alpha}\delta^{(3)}({\bf x} -{\bf x}_0)\bigr).
\label{eq:poly2axial}
\end{eqnarray}
We see that the equation (\ref{eq:polyexpaxial2}) is obtained after the integration ${\cal D}\Pi_A^{ak}$.
Let us note that the first term in the fourth line of Eq.(\ref{eq:poly2axial}) is rewritten by Eq.(\ref{eq:adot2axial}).
If we expand the exponent of the Gaussian integral (\ref{eq:polyatauaxial}), we have
\begin{eqnarray}
&&-\half 
\Bigl(iA_{\tau}^a -({\cal D}^{-1})^{ac}\bigl(i (D_k\dot{A}_k)^c +J_F^{c0}  +
g ({\tilde T}^c)_{\alpha\alpha}\delta^{(3)}({\bf x}-{\bf x}_0) \bigr)\Bigr)
\nonumber\\
&&\times ({\cal D})^{ad}
\Bigl(iA_{\tau}^d -({\cal D}^{-1})^{de}\bigl(i (D_l\dot{A}_l)^e +J_F^{e0}  + 
g ({\tilde T}^e)_{\alpha\alpha}\delta^{(3)}({\bf x}-{\bf x}_0) \bigr)\Bigr)
\nonumber\\
&=&-\half  (D_k A_{\tau})^a(D_k A_{\tau})^a 
-\half \del_3A_{\tau}^a \del_3A_{\tau}^a + iA_{\tau}^a J_F^{a0} +\dot{A}_k^a(D_kA_{\tau})^a
\nonumber\\
&&+\half (D_k\dot{A}_k)^a({\cal D}^{-1}D_l\dot{A}_l)^a
+i \dot{A}_k^a(D_k{\cal D}^{-1}J_F^0)^a -\half J_F^{a0}({\cal D}^{-1}J_F^0)^a
\nonumber\\
&&+ig A_{\tau}^a ({\tilde T}^a)_{\alpha\alpha}\delta^{(3)}({\bf x}-{\bf x}_0)\nonumber\\
&&
-g({\tilde T}^a)_{\alpha\alpha}\delta^{(3)}({\bf x} -{\bf x}_0)({\cal D}^{-1}J_F^0)^a 
-ig ({\tilde T}^a)_{\alpha\alpha}\delta^{(3)}({\bf x} -{\bf x}_0)({\cal D}^{-1}D_k\dot{A}_k)^a
\nonumber\\
&&
-\frac{g^2}{2} \bigl(({\tilde T}^a)_{\alpha\alpha}\delta^{(3)}({\bf x} -{\bf x}_0)\bigr)
\bigl(
({\cal D}^{-1})^{ab}({\tilde T}^b)_{\alpha\alpha}\delta^{(3)}({\bf x} -{\bf x}_0)
\bigr).
\label{eq:atauexpan}
\end{eqnarray}
We find that some of the terms in Eqs.(\ref{eq:poly2axial}) and (\ref{eq:atauexpan}) are canceled each 
other, and we obtain the result (\ref{eq:axialpoly}) in the text. The fifth 
line in Eq.(\ref{eq:atauexpan}) corresponds to the Polyakov loop as it should. Let 
us note that the last term in Eq.(\ref{eq:atauexpan}) is
canceled with the last term in Eq.(\ref{eq:poly2axial}). As for the term with the two delta functions, only the 
seventh line in Eq.(\ref{eq:poly2axial}) is left, which is the divergent self-energy of the point color charge 
densities owing to introducing the operator (\ref{eq:polyopdefaxial}) as discussed in the text.
\section{Coulomb gauge}
\label{sec:canonicalcoul}
Let us present the formulae and relations in order to obtain the results of the section
\ref{sec:secfive} in the text.
\subsection{Hamiltonian in Coulomb gauge}
\label{sec:canohamcoul}
Let us recall the operators defined in Eqs.(\ref{eq:defc}) and (\ref{eq:deftildec}), from which we obtain
new operators given by Eqs.(\ref{eq:ctildeprime}) and (\ref{eq:cprime}). These are proved by using
$\del_l^2 =\Delta - (\del_3)^2$ like
\begin{eqnarray}
({\tilde C}^{\prime}_k)^{ab} 
&\equiv & (\delta_{kl} - \del_k\Delta^{-1}\del_l)({\tilde C}_l)^{ab} 
=(\delta_{kl} - \del_k\Delta^{-1}\del_l)\bigl((D_l)^{ab} -\del_l(\del_3)^{-1}(D_3)^{ab}\bigr)\nonumber\\
&=&(D_k)^{ab} -\del_k\Delta^{-1}\del \cdot (D)^{ab} ,
\label{eq:ctildeprimeapp}
\\
(C^{\prime}_k)^{ab} 
&\equiv & (C_l)^{ab}(\delta_{lk} - \del_l\Delta^{-1}\del_k)
=\bigl((D_l)^{ab} -(D_3)^{ab}(\del_3)^{-1}\del_l\bigr)(\delta_{lk} - \del_l\Delta^{-1}\del_k)
\nonumber\\
&=&(D_k)^{ab} -(D)^{ab}\cdot\del \Delta^{-1}\del_k,
\label{eq:cprimeapp}
\end{eqnarray}
where we have used the notation $\del_i(D_i)^{ab}=\del\cdot (D)^{ab}$ which is 
also $(D)^{ab}\cdot\del$ due to the Coulomb gauge, $\del_iA^{ai}=0$.

We also find the useful formulae given by
\begin{eqnarray}
(C_k{\tilde C}_k^{\prime})^{ab} &=&
\bigl((D_k)^{ac} -(D_3)^{ac}(\del_3)^{-1}\del_k\bigr)
\bigl((D_k)^{cb} -\del_k\Delta^{-1}\del \cdot (D)^{cb}\bigr)\nonumber\\
&=&(D_k^2)^{ab} + (D_3^2)^{ab} -(D_j\del_j{\Delta}^{-1}\del_iD_i)^{ab} \nonumber\\
&=&{\cal D}^{ab} - ({\tilde \Delta})^{ab},
\label{eq:ctildecrel1app}\\
(C_k^{\prime}{\tilde C}_k)^{ab} &=&
\bigl((D_k)^{ac} -(D)^{ac}\cdot\del\Delta^{-1}\del_k\bigr)
\bigl((D_k)^{cb} -\del_l(\del_3)^{-1}(D_3)^{ac}\bigr)
\nonumber\\
&=&(D_k^2)^{ab} + (D_3^2)^{ab} - (D_i\del_i{\Delta}^{-1}\del_jD_j)^{ab} \nonumber\\
&=&{\cal D}^{ab} - ({\tilde \Delta})^{ab},
\label{eq:ctildecrel2app}
\end{eqnarray}
where ${\tilde \Delta}$ is defined by $D_i\del_i{\Delta}^{-1}\del_jD_j$ whose inverse operator is given
by Eq.(\ref{eq:tildedelta}).

The Lagrangian ${\cal L}(A_k^a, \psi)_{\rm Coul}$ is obtained by imposing the Coulomb gauge $\del_i A^{ai}=0$ 
and eliminating $A_0^a$ by using the constraint (\ref{eq:constcoul}) in the Lagrangian (\ref{eq:lag}), and 
after straightforward calculations we have
\begin{eqnarray}
{\cal L}(A_k^a, \psi)_{\rm Coul}&=&
\half \dot{A}_k^a(N_{kl}\dot{A}_l)^a 
-\dot{A}_k^a({\tilde C}_k{\cal D}^{-1} J_F^0)^a
+\half J_F^{a0}({\cal D}^{-1}J_F^0)^a
\nonumber\\
&&-\frac{1}{4}F_{kl}^aF_{kl}^a -\half F_{3k}^aF_{3k}^a 
+{\bar\psi}i\gamma^0\dot{\psi} +{\bar\psi}i\gamma^kD_k\psi +{\bar\psi}i\gamma^3D_3\psi ,
\label{eq:tildelagcoul}
\end{eqnarray}
where we have used the definition (\ref{eq:ndefcoul}). In 
deriving (\ref{eq:tildelagcoul}), we have formally performed the partial integration with
respect to ${\tilde C}_k$ and $\cal D$. The partial integration with respect to
${\tilde C}_k$ gives a new operator $C_k$, accompanying a minus sign and vice versa. 
Incidentally, the partial integration with respect to $(N)_{kl}^{ab}, (N^{-1})_{kl}^{ab}$ does not accompany a 
minus sign.

Let us show that $(N^{-1})_{kl}^{ab}$ is given by Eq.(\ref{eq:ninverse}). To this end, we
first introduce a new operator defined by
\begin{equation}
(N')_{km}^{ab}\equiv (\delta_{kl} -\del_k\Delta^{-1}\del_l)(N)_{lm}^{ab} .
\label{eq:nprime}
\end{equation}
$(N)_{km}^{ab}$ is defined by Eq.(\ref{eq:ndefcoul}) in the text. Since it is easy to show that 
\begin{equation}
(\delta_{kl} -\del_k\Delta^{-1}\del_l)(\delta_{lm} +\del_l(\del_3)^{-2}\del_m)=\delta_{km},
\label{eq:identitycoulapp}
\end{equation}
the equation (\ref{eq:nprime}) becomes 
\begin{equation}
(N')_{km}^{ab}=\delta_{km}\delta^{ab} -({\tilde C}_k^{\prime} {\cal D}^{-1}C_m)^{ab},
\label{eq:nprimeexp}
\end{equation}
where we have used  the definition (\ref{eq:ctildeprimeapp}).
From Eq.(\ref{eq:nprime}), we obtain that 
\begin{equation}
(N^{-1})_{kl}^{ab} =(N^{-1\prime})_{km}^{ab}(\delta_{ml} -\del_m\Delta^{-1}\del_l).
\label{eq:ninvnprime}
\end{equation}
The explicit form of $({N^{-1}}^{\prime})^{ab}_{kl}$ is easily found to be
\begin{equation}
({N^{-1}}^{\prime})_{kl}^{ab}=\del_{kl}\delta^{ab} + ({\tilde C}^{\prime}_k{\tilde \Delta}^{-1}C_l)^{ab}.
\label{eq:nprimeinvexp}
\end{equation}
This is, by using the relation (\ref{eq:ctildecrel1app}), because
\begin{eqnarray}
&&(N')_{kl}^{ac}(N^{-1\prime })_{lm}^{cb} \nonumber\\
&=&
\bigl(\delta_{kl}\delta^{ac} -({\tilde C}_k^{\prime} {\cal D}^{-1}C_l)^{ac}\bigr)
\bigl(\del_{lm}\delta^{cb} + ({\tilde C}^{\prime}_l{\tilde \Delta}^{-1}C_m)^{cb}\bigr) \nonumber\\
&=&\delta_{km}\delta^{ab} +({\tilde C}^{\prime}_k{\tilde\Delta}^{-1}C_m)^{ab}
-({\tilde C}_k^{\prime}{\cal D}^{-1}C_m)^{ab}
-({\tilde C}_k^{\prime}{\cal D}^{-1}C_l{\tilde C}_l^{\prime}{\tilde \Delta}^{-1}C_m)^{ab}\nonumber\\
&=&
\delta_{km}\delta^{ab} +({\tilde C}^{\prime}_k{\tilde\Delta}^{-1}C_m)^{ab}
-({\tilde C}_k^{\prime}{\cal D}^{-1}C_m)^{ab}
-({\tilde C}_k^{\prime}{\cal D}^{-1}({\cal D} - {\tilde \Delta}) {\tilde \Delta}^{-1}C_m)^{ab}\nonumber\\
&=&\delta_{km}\delta^{ab} .
\end{eqnarray}
Then, the equation (\ref{eq:ninvnprime}) is evaluated as
\begin{eqnarray}
(N^{-1})_{kl}^{ab}&=&
\bigl(\delta_{km}\delta^{ab} + ({\tilde C}_k^{\prime}{\tilde \Delta}^{-1}C_m)^{ab}\bigr)(\delta_{ml} - \del_m \Delta^{-1}\del_l)
\nonumber\\
&=&(\delta_{kl} -\del_k\Delta^{-1}\del_l)\delta^{ab} +({\tilde C}_k^{\prime}{\tilde \Delta}^{-1}C_l^{\prime})^{ab} ,
\label{eq:ninverseapp}
\end{eqnarray}
where we have used the definition (\ref{eq:cprimeapp}) for $C_l^{\prime}$. We have 
finished to prove the equation (\ref{eq:ninverse}).

As in the case for the axial gauge, the Hamiltonian (\ref{eq:hamiltoniancoul}), 
\begin{eqnarray}
{\cal H}_{\rm Coul} &=& \Pi_A^{ak}\dot{A}_k^a +\Pi_{\psi}\dot{\psi} - {{\cal L}}(A_k, \psi)_{\rm Coul}
\nonumber\\
&=&
\half \Pi_A^{ak}(N^{-1}_{kl}\Pi_A^l)^a + \Pi_A^{ak}(N^{-1}_{kl}{\tilde C}_l{\cal D}^{-1}J_F^0)^a
\nonumber\\
&&+\half ({\tilde C}_k{\cal D}^{-1}J_F^0)(N^{-1}_{kl}{\tilde C}_l{\cal D}^{-1}J_F^0)^a 
-\half J_F^{a0}({\cal D}^{-1}J_F^0)^a
\nonumber\\
&&+\frac{1}{4}F_{kl}^a F_{kl}^a +\half F_{3k}^aF_{3k}^a - \bar\psi i \gamma^k D_k \psi 
-\bar\psi i \gamma^3D_3\psi 
\label{eq:hamiltoniancoulapp}
\end{eqnarray}
can be rewritten into another form. The second and the third terms in Eq.(\ref{eq:hamiltoniancoulapp}) 
after the partial integration with respect to $N_{kl}^{-1}, {\tilde C}_k$ and ${\cal D}^{-1}$ are rewritten as
\begin{eqnarray}
&&
-({\cal D}^{-1}C_lN^{-1}_{lk}\Pi_A^{k})^aJ_F^{a0}
-\half J_F^{0a}({\cal D}^{-1}C_kN^{-1}_{kl}{\tilde C}_l{\cal D}^{-1}J_F^0)^a \nonumber\\
&&=\Pi_A^{ak}\bigl({\tilde C}_k^{\prime}{\tilde \Delta}^{-1}J_F^0\bigr)^a 
-\half J_F^{a0}\Big(\bigl(\bigl({\tilde \Delta}^{-1}\bigr)^{ab} - ({\cal D}^{-1})^{ab}\bigr)J_F^{b0}\Bigr),
\label{eq:relcoulapp}
\end{eqnarray}
where we have used the relations,
\begin{eqnarray}
({\cal D}^{-1}C_lN^{-1}_{lk})^{ab}=({\tilde \Delta}^{-1}C_k^{\prime})^{ab}
%
%
%
\label{eq:newrelcoul1}
\end{eqnarray}
and 
\begin{eqnarray}
\Bigl({\cal D}^{-1}C_kN_{kl}^{-1}{\tilde C}_l{\cal D}^{-1}\Bigr)^{ab}=({\tilde \Delta}^{-1})^{ab} -({\cal D}^{-1})^{ab}.
%
%
%
\label{eq:newrelcoul2}
\end{eqnarray}
The above relations can be shown by using the equations (\ref{eq:ctildecrel1app}), (\ref{eq:ctildecrel2app})
and the explicit form of $(N^{-1})_{kl}^{ab}$ given by Eq.(\ref{eq:ninverseapp}). We 
have finally performed the partial integration with respect to ${\tilde\Delta}^{-1}$ and $C_k^{\prime}$
in order to arrive at Eq.(\ref{eq:relcoulapp}). Note that the partial integration with respect to
$C_k^{\prime}$ yields a new operator ${\tilde C}_k^{\prime}$, accompanying a minus sign and vice versa, while
the partial integration of ${\tilde \Delta}^{-1}$ does not have a minus sign. Then, we obtain that
\begin{eqnarray}
{\cal H}_{\rm Coul} &=&
\half \Pi_A^{ak}(N^{-1}_{kl}\Pi_A^l)^a + \Pi_A^{ak}({\tilde C}^{\prime}_k{\tilde \Delta}^{-1}J_F^0)^a
-\half J_F^{a0}({\tilde \Delta}^{-1}J_F^{0})^a\nonumber\\
&&+\frac{1}{4}F_{kl}^a F_{kl}^a +\half F_{3k}^aF_{3k}^a - \bar\psi i \gamma^k D_k \psi 
-\bar\psi i \gamma^3D_3\psi .
\label{eq:anohamcoul}
\end{eqnarray}
Let us note that the third term in the first line of Eq.(\ref{eq:anohamcoul}) is well-known to be the self-energy
of the point color charge densities of fermion in the Coulomb gauge. 
Unlike the case for the axial gauge, it depends on the gauge field through the operator ${\tilde \Delta}^{-1}$.
We also point out the correspondence of the Hamiltonians (\ref{eq:anoham}) and (\ref{eq:anohamcoul})
such as $M^{-1}\leftrightarrow N^{-1}, D_k\leftrightarrow {\tilde C}_k^{\prime}$ 
and $(\del_3)^{-2} \leftrightarrow {\tilde\Delta}^{-1}$, though the explicit forms for the operators are 
quite different.
\subsection{Trace formula in Coulomb gauge}
\label{sec:tracecoul}
We complete the square with respect to $\Pi_A^{ak}$ in the 
exponent of Eq.(\ref{eq:coulexp}) in order to perform the integration ${\cal D}\Pi_A^{ak}$, 
\begin{eqnarray}
&&i\Pi_A^{ak}\dot{A}_k^a +i \Pi_{\psi}\dot{\psi} -{\cal H}_{\rm Coul}\nonumber\\
&=&-\half 
\Bigl(\Pi_A^{ak} -i (N_{kl}\dot{A}_l)^a +({\tilde C}_k {\cal D}^{-1}J_F^0)^a\Bigr)
(N^{-1})_{km}^{ac}\Bigl(\Pi_A^{cm} -i (N_{mn}\dot{A}_n)^c +({\tilde C}_m {\cal D}^{-1}J_F^0)^c\Bigr)
\nonumber\\
&&
-\half \dot{A}_k^a(N_{kl}\dot{A}_l)^a -i\dot{A}_k^a({\tilde C}_k{\cal D}^{-1}J_F^0)^a 
+\half J_F^{a0}({\cal D}^{-1}J_F^0)^a 
-\frac{1}{4}F_{kl}^a F_{kl}^a -\half F_{3k}^aF_{3k}^a \nonumber\\
&&
+i\Pi_{\psi}\dot{\psi}+ \bar\psi i \gamma^k D_k \psi  +\bar\psi i \gamma^3D_3\psi .
\label{eq:quadracoul}
\end{eqnarray}
The first term in the third line of Eq.(\ref{eq:quadracoul}) is rewritten as
\begin{equation}
-\half \dot{A}_k^a(N_{kl}\dot{A}_l)^a=
-\half \dot{A}_k^a\dot{A}_k^a 
-\half  \bigl((\del_3)^{-1}\del_k\dot{A}_k^a\bigr)\bigl(  (\del_3)^{-1}\del_l\dot{A}_l^a\bigr)
+\half \dot{A}_k^a({\tilde C}_k{\cal D}^{-1}C_l\dot{A}_l)^a ,
\label{eq:adot2n}
\end{equation}
where we have used the explicit form
for $(N)_{kl}^{ab}$, Eq.(\ref{eq:ndefcoul}) and performed the partial integration with 
respect to $(\del_3)^{-1}$ and $\del_k$.

The expansion of the exponent of the Gaussian integral (\ref{eq:gausscoul}) yields
\begin{eqnarray}
&&-\half 
\Bigl(iA_{\tau}^a -({\cal D}^{-1})^{ac}\bigl(i (C_k\dot{A}_k)^c +J_F^{c0}  \bigr)\Bigr)
 ({\cal D})^{ad}
\Bigl(iA_{\tau}^d -({\cal D}^{-1})^{de}\bigl(i (C_l\dot{A}_l)^e +J_F^{e0}   \bigr)\Bigr)
\nonumber\\
&=&-\half  (D_i A_{\tau})^a(D_i A_{\tau})^a + iA_{\tau}^a J_F^{a0} + \dot{A}_k^a(D_kA_{\tau})^a
-\half \dot{A}_l^a({\tilde C}_l{\cal D}^{-1}C_k\dot{A}_k)^a
\nonumber\\
&&
+i \dot{A}_k^a({\tilde C}_k{\cal D}^{-1}J_F^0)^a -\half J_F^{a0}({\cal D}^{-1}J_F^0)^a
\label{eq:ataucoul2}
\end{eqnarray}
Let us note that the third and the fourth terms in Eq.(\ref{eq:ataucoul2}) comes from one of
the terms obtained by expanding the exponent of Eq.(\ref{eq:ataucoul2}),
\begin{equation}
-A_{\tau}^a\bigl(C_k\dot{A}_k\bigr)^a = \bigl(D_kA_{\tau}\bigr)^a\dot{A}_k^a -
\bigl(D_3A_{\tau}\bigr)^a\bigl((\del_3)^{-1}\del_k\dot{A}_k^a\bigr),
\end{equation}
where we have used the equation (\ref{eq:defc}) and the partial integration with respect to $D_k$ has been done. 
We see that some of the terms in Eqs. (\ref{eq:quadracoul}) and (\ref{eq:ataucoul2}) are canceled each 
other to yield the result (\ref{eq:coulatau}) in the text.
\subsection{Polyakov loop in Coulomb gauge}
\label{sec:polycoul}
%
%
%
%
%
%
%
We need to complete the square with respect to $\Pi_A^{ak}$ in the exponent of Eq.(\ref{eq:polyoexp})
in order to perform the integration ${\cal D}\Pi_A^{ak}$. It is given by
\begin{eqnarray}
&&-\half \Bigl(\Pi_A^{ak} - (N)_{kl}^{ab}\bigl(i\dot{A}_l^b  - 
g({\tilde C}_l^{\prime}{\tilde \Delta}^{-1})^{bd}({\tilde T}^d)_{\alpha\alpha}\delta^{(3)}({\bf x} -{\bf x}_0)\bigr)
+({\tilde C}_k {\cal D}^{-1}J_F^0)^b\Bigr)(N^{-1})_{km}^{ag}
\nonumber\\
&&\times 
\Bigl(\Pi_A^{gm} - (N)_{mn}^{gc}\bigl(i\dot{A}_n^c 
- g({\tilde C}_n^{\prime}{\tilde \Delta}^{-1})^{cf}({\tilde T}^f)_{\alpha\alpha}\delta^{(3)}({\bf x} -{\bf x}_0)\bigr)
+({\tilde C}_m {\cal D}^{-1}J_F^0)^g\Bigr).
\end{eqnarray}
There appear new terms with the gauge coupling $g$ because of introducing the operator
(\ref{eq:coulodef}), which we do not have in Eq.(\ref{eq:quadracoul}). According to the perfect square 
with respect to $\Pi_A^{ak}$, we should add the following terms, 
\begin{eqnarray}
&&-g({\tilde T}^a)_{\alpha\alpha}\delta^{(3)}({\bf x} -{\bf x}_0)\bigl({\tilde \Delta}^{-1}C_k^{\prime}\Pi_A^k\bigr)^a
+ig ({\tilde T}^a)_{\alpha\alpha}\delta^{(3)}({\bf x} -{\bf x}_0)({\tilde\Delta}^{-1}C_k'N_{kl}\dot{A}_l)^a\nonumber\\
&&+g(N)_{kl}^{ab}({\tilde C}_l^{\prime}{\tilde \Delta}^{-1})^{bd}({\tilde T}^d)_{\alpha\alpha}\delta^{(3)}({\bf x}-{\bf x}_0)
\bigl((N^{-1})_{km}^{ag}({\tilde C}_k{\cal D}^{-1}J_F^0)^g\bigr)
\nonumber\\
&&-\frac{g^2}{2} \bigl(({\tilde T}^a)_{\alpha\alpha}(\del_3)^{-2}\delta^{(3)}({\bf x} -{\bf x}_0)\bigr)
\bigl(
({\tilde \Delta}^{-1}C_k^{\prime}N_{kl}{\tilde C}_l^{\prime}
{\tilde \Delta}^{-1})^{ab}({\tilde T}^b)_{\alpha\alpha}\delta^{(3)}({\bf x} -{\bf x}_0)
\bigr),
\end{eqnarray}
where we have performed the partial integration with respect 
to ${\tilde \Delta}^{-1}, {\tilde C}_k^{\prime}$ and $(N)_{kl}^{ab}$. 
They can be rewritten into simple forms by using the explicit form of $N_{kl}^{ab}$.
With the help of Eqs.(\ref{eq:cprimeapp}), (\ref{eq:ctildecrel1app}), (\ref{eq:ctildecrel2app}) 
and (\ref{eq:identitycoulapp}) we obtain that
\begin{eqnarray}
({\tilde \Delta}^{-1}C_k^{\prime}N_{kl})^{ab}&=&
({\tilde \Delta}^{-1}C_k^{\prime})^{ac}
\bigl( \delta^{cb}(\delta_{kl}+\del_k(\del_3)^{-1}\del_l) -({\tilde C}_k{\cal D}^{-1}C_l)^{cb}\bigr)\nonumber\\
&=&({\tilde \Delta}^{-1})^{ac}(C_l)^{cb}-({\tilde \Delta}^{-1})^{ac}(C_k^{\prime}{\tilde C}_k)^{cd}({\cal D}^{-1}C_l)^{db}
\nonumber\\
&=&({\tilde \Delta}^{-1})^{ac}(C_l)^{cb}-({\tilde \Delta}^{-1})^{ac}({\cal D} -{\tilde \Delta})^{cd}({\cal D}^{-1}C_l)^{db}
\nonumber\\
&=&({\cal D}^{-1}C_l)^{ab}
\label{eq:coulnewrel1}
\end{eqnarray}
and
\begin{eqnarray}
({\tilde \Delta}^{-1}C_k^{\prime}N_{kl}{\tilde C}_l^{\prime}{\tilde \Delta}^{-1})^{ab}
&=&({\tilde \Delta}^{-1}C_k^{\prime}N_{kl})^{ac}({\tilde C}_l^{\prime}{\tilde \Delta}^{-1})^{cb}
\stackrel{\rm(\ref{eq:coulnewrel1})}{=}({\cal D}^{-1}C_l)^{ac}({\tilde C}_l^{\prime}{\tilde \Delta}^{-1})^{cb}
\nonumber\\
&=&({\cal D}^{-1})^{ac}(C_l{\tilde C}_l^{\prime})^{cd}({\tilde\Delta}^{-1})^{db}
=({\cal D}^{-1})^{ac}({\cal D}-{\tilde \Delta})^{cd}({\tilde\Delta}^{-1})^{db}
\nonumber\\
&=&({\tilde \Delta}^{-1})^{ab}-({\cal D}^{-1})^{ab}.
\label{eq:coulnewrel2}
\end{eqnarray}
Then, we have
\begin{eqnarray}
&&-g({\tilde T}^a)_{\alpha\alpha}\delta^{(3)}({\bf x} -{\bf x}_0)\bigl({\tilde \Delta}^{-1}C_k^{\prime}\Pi_A^k\bigr)^a
+ig ({\tilde T}^a)_{\alpha\alpha}\delta^{(3)}({\bf x} -{\bf x}_0)({\tilde\Delta}^{-1}C_k'N_{kl}\dot{A}_l)^a\nonumber\\
&&
+g(N)_{kl}^{ab}({\tilde C}_l^{\prime}{\tilde \Delta}^{-1})^{bd}({\tilde T}^d)_{\alpha\alpha}\delta^{(3)}({\bf x}-{\bf x}_0)
\bigl((N^{-1})_{km}^{ag}({\tilde C}_k{\cal D}^{-1}J_F^0)^g)\bigr)
\nonumber\\
&&-\frac{g^2}{2} \bigl(({\tilde T}^a)_{\alpha\alpha}(\del_3)^{-2}\delta^{(3)}({\bf x} -{\bf x}_0)\bigr)
\bigl(
({\tilde \Delta}^{-1}C_k^{\prime}N_{kl}{\tilde C}_l^{\prime}
{\tilde \Delta}^{-1})^{ab}({\tilde T}^b)_{\alpha\alpha}\delta^{(3)}({\bf x} -{\bf x}_0)
\bigr)\nonumber\\
&=&
-g({\tilde T}^a)_{\alpha\alpha}\delta^{(3)}({\bf x} -{\bf x}_0)\Bigl(({\tilde \Delta}^{-1})^{ab}
\bigl((C_k^{\prime}\Pi_A^k)^a+J_F^{0b}\bigr)\Bigr)
\nonumber\\
&&
+g({\tilde T}^a)_{\alpha\alpha}\delta^{(3)}({\bf x} -{\bf x}_0)({\cal D}^{-1}J_F^0)^a 
+ig({\tilde T}^a)_{\alpha\alpha}\delta^{(3)}({\bf x} -{\bf x}_0)({\cal D}^{-1}C_k\dot{A}_k^a)\nonumber\\
&&
-\frac{g^2}{2}({\tilde T}^a)_{\alpha\alpha}\delta^{(3)}({\bf x} -{\bf x}_0)
\bigl(({\tilde \Delta}^{-1})^{ab}({\tilde T}^b)_{\alpha\alpha}\delta^{(3)}({\bf x} -{\bf x}_0)\bigr)
\nonumber\\
&&+\frac{g^2}{2}({\tilde T}^a)_{\alpha\alpha}\delta^{(3)}({\bf x} -{\bf x}_0)
\bigl(({\cal D}^{-1})^{ab}({\tilde T}^b)_{\alpha\alpha}\delta^{(3)}({\bf x} -{\bf x}_0)\bigr),
\label{eq:coulnewrel3}
\end{eqnarray}
where the first term, aside from the minus sign, in the right-hand-side is the exponent of operator
(\ref{eq:coulodef}). Hence, the exponent of the right-hand-side of Eq.(\ref{eq:polyoexp}) is written 
into the quadratic form with respect to $\Pi_A^{ak}$ as
\begin{eqnarray}
&&
g({\tilde T}^a)_{\alpha\alpha}\delta^{(3)}({\bf x}-{\bf x}_0)
\Bigl(
({\tilde \Delta}^{-1})^{ab}\bigl((C_k^{\prime}\Pi_A^k)^b +J_F^{b0}+ \delta_{\rm self}^b\bigr)
\Bigr) +
i\Pi_A^{ak}\dot{A}_k^a +i \Pi_{\psi}\dot{\psi} -{\cal H}_{\rm Coul}\nonumber\\
&=&-\half \Bigl(\Pi_A^{ak} - (N)_{kl}^{ab}\bigl(i\dot{A}_l^b  - 
g({\tilde C}_l^{\prime}{\tilde \Delta}^{-1})^{bd}({\tilde T}^d)_{\alpha\alpha}\delta^{(3)}({\bf x} -{\bf x}_0)\bigr)
+({\tilde C}_k {\cal D}^{-1}J_F^0)^b\Bigr)(N^{-1})_{km}^{ag}
\nonumber\\
&&\times 
\Bigl(\Pi_A^{gm} - (N)_{mn}^{gc}\bigl(i\dot{A}_n^c 
- g({\tilde C}_n^{\prime}{\tilde \Delta}^{-1})^{cf}({\tilde T}^f)_{\alpha\alpha}\delta^{(3)}({\bf x} -{\bf x}_0)\bigr)
+({\tilde C}_m {\cal D}^{-1}J_F^0)^g\Bigr)
\nonumber\\
&&
-\half \dot{A}_k^a(N_{kl}\dot{A}_l)^a -i\dot{A}_k^a({\tilde C}_k{\cal D}^{-1}J_F^0)^a 
+\half J_F^{a0}({\cal D}^{-1}J_F^0)^a 
-\frac{1}{4}F_{kl}^a F_{kl}^a -\half F_{3k}^aF_{3k}^a \nonumber\\
&&
+i\Pi_{\psi}\dot{\psi}+ \bar\psi i \gamma^k D_k \psi  +\bar\psi i \gamma^3D_3\psi 
\nonumber\\
&&
+g({\tilde T}^a)_{\alpha\alpha}\delta^{(3)}({\bf x} -{\bf x}_0)
({\tilde \Delta}^{-1})^{ab}\delta_{\rm self}^b\nonumber\\
&&
+g({\tilde T}^a)_{\alpha\alpha}\delta^{(3)}({\bf x} -{\bf x}_0) ({\cal D}^{-1}J_F^0)^a 
+ig({\tilde T}^a)_{\alpha\alpha}\delta^{(3)}({\bf x} -{\bf x}_0)({\cal D}^{-1}C_k\dot{A}_k)^a
\nonumber\\
&&
-\frac{g^2}{2}({\tilde T}^a)_{\alpha\alpha}\delta^{(3)}({\bf x} -{\bf x}_0)
\bigl(({\tilde \Delta}^{-1})^{ab}({\tilde T}^b)_{\alpha\alpha}\delta^{(3)}({\bf x} -{\bf x}_0)\bigr)
\nonumber\\
&&+\frac{g^2}{2}({\tilde T}^a)_{\alpha\alpha}\delta^{(3)}({\bf x} -{\bf x}_0)
\bigl(({\cal D}^{-1})^{ab}({\tilde T}^b)_{\alpha\alpha}\delta^{(3)}({\bf x} -{\bf x}_0)\bigr).
\label{eq:poly2coul}
\end{eqnarray}
We observe that the integration ${\cal D}\Pi_A^{ak}$ yields the equation (\ref{eq:polyexpcoul2}).
Let us note that the first term in the fourth line of Eq.(\ref{eq:poly2coul}) is rewritten by Eq.(\ref{eq:adot2n}).
As explained in the text, the counter term, the sixth line of Eq.(\ref{eq:poly2coul}), exactly cancels the
divergent self-energy, the eighth line of Eq.Eq.(\ref{eq:poly2coul}).

If we expand the exponent of the Gaussian integral (\ref{eq:polyataucoul}), we have
\begin{eqnarray}
&&-\half 
\Bigl(iA_{\tau}^a -({\cal D}^{-1})^{ac}\bigl(i (C_k\dot{A}_k)^c +J_F^{c0}  +g ({\tilde T}^c)_{\alpha\alpha}
\delta^{(3)}({\bf x}-{\bf x}_0) \bigr)\Bigr)
\nonumber\\
&&~\times ({\cal D})^{ad}
\Bigl(iA_{\tau}^d -({\cal D}^{-1})^{de}\bigl(i (C_l\dot{A}_l)^e +J_F^{e0}  + g ({\tilde T}^e)_{\alpha\alpha}
\delta^{(3)}({\bf x}-{\bf x}_0) \bigr)\Bigr)
\nonumber\\
&=&-\half  (D_i A_{\tau})^a(D_i A_{\tau})^a + iA_{\tau}^a J_F^{a0} -A_{\tau}^a (C_k\dot{A}_k)^a
\nonumber\\
&&-\half \dot{A}_l^a({\tilde C}_l{\cal D}^{-1}C_k\dot{A}_k)^a
+i \dot{A}_k^a({\tilde C}_k{\cal D}^{-1}J_F^0)^a -\half J_F^{a0}({\cal D}^{-1}J_F^0)^a
\nonumber\\
&&+ig A_{\tau}^a ({\tilde T}^a)_{\alpha\alpha}\delta^{(3)}({\bf x}-{\bf x}_0)
\nonumber\\
&&-g({\tilde T}^a)_{\alpha\alpha}\delta^{(3)}({\bf x} -{\bf x}_0)({\cal D}^{-1}J_F^0)^a 
-ig ({\tilde T}^a)_{\alpha\alpha}\delta^{(3)}({\bf x} -{\bf x}_0)({\cal D}^{-1}C_k\dot{A}_k)^a
\nonumber\\
&&
-\frac{g^2}{2} ({\tilde T}^a)_{\alpha\alpha}\delta^{(3)}({\bf x} -{\bf x}_0)
\bigl(
({\cal D}^{-1})^{ab}({\tilde T}^b)_{\alpha\alpha}\delta^{(3)}({\bf x} -{\bf x}_0)
\bigr).
\label{eq:polyataucoul2}
\end{eqnarray}
The third term in the first line of the 
right-hand-side of Eq.(\ref{eq:polyataucoul2}) can be written by 
\begin{equation}
-A_{\tau}^a (C_k\dot{A}_k)^a= \dot{A}_k^a(D_kA_{\tau})^a - (D_3A_{\tau})^a((\del_3)^{-1}\del_k\dot{A}_k^a),
\label{eq:polyrel3}
\end{equation}
where we have performed the partial integration with respect to $D_3$. We observe that some 
of the terms in Eqs.(\ref{eq:poly2coul}) and (\ref{eq:polyataucoul2}) are 
canceled each other to result the equation (\ref{eq:coulpoly}) in the text.
\section{Evaluation of Determinants}
\subsection{Axial gauge}
\label{sec:detaxial}
Let us prove the equation (\ref{eq:determinantaxial}). We evaluate ${\rm det }(M)$ where $M$ is 
defined by Eq.(\ref{eq:mexplicit}). To this end, we write $M$ as
\begin{equation}
(M)_{kl}^{ab}=\begin{pmatrix}
\delta^{ab} -(D_1{\cal D}^{-1}D_1)^{ab} & -(D_1{\cal D}^{-1}D_2)^{ab} \\
  -(D_2{\cal D}^{-1}D_1)^{ab}& \delta^{ab} -(D_2{\cal D}^{-1}D_2)^{ab}\\
\end{pmatrix}\qquad (a, b=1\sim N^2-1).
\end{equation}
Introducing the relation given by 
\begin{equation}
{\rm det}~
\begin{pmatrix}
0 & \mathbb{I}_{N^2-1} \\
\mathbb{I}_{N^2-1} & 0\\
\end{pmatrix}=(-1)^{N^2-1},
\end{equation}
where $\mathbb{I}_{N^2-1}$ is the $(N^2 -1)\times (N^2 -1)$ unit matrix, we have
\begin{eqnarray}
(-1)^{N^2-1}~{\rm det}(M)&=&
{\rm det}\Biggl[\begin{pmatrix}
0 & \mathbb{I}_{N^2 -1} \\
\mathbb{I}_{N^2-1} & 0\\
\end{pmatrix}M\Biggr]\nonumber\\
&=&{\rm det}
\begin{pmatrix}
-D_2{\cal D}D_1 & \mathbb{I}_{N^2-1} -D_2{\cal D}^{-1}D_2 \\
\mathbb{I}_{N^2-1} -D_1{\cal D}^{-1}D_1 & -D_1{\cal D}^{-1}D_2\\
\end{pmatrix}.
\label{eq:relationaxial}
\end{eqnarray}
With help of the formula,
\begin{equation}
{\rm det}\begin{pmatrix}
A & C \\
D & B\\
\end{pmatrix}
={\rm det}
\begin{pmatrix}
A & 0 \\
D & \mathbb{I}\\
\end{pmatrix}
\begin{pmatrix}
\mathbb{I}& A^{-1}C \\
0 & B-DA^{-1}C\\
\end{pmatrix}
={\rm det}(A)\times {\rm det}(B-DA^{-1}C),
\label{eq:matformula}
\end{equation} 
The equation (\ref{eq:relationaxial}) becomes
\begin{eqnarray}
&&{\rm det}(-D_2{\cal D}D_1){\rm det}\Bigl(-D_1{\cal D}^{-1}D_2 -
(\mathbb{I}_{N^2-1} -D_1{\cal D}^{-1}D_1)(-D_2{\cal D}^{-1}D_1)^{-1}(\mathbb{I}_{N^2-1} -D_2{\cal D}^{-1}D_2)\Bigr)
\nonumber\\
&=&{\rm det}\Bigl(D_2{\cal D}^{-1}(D_1)^2{\cal D}^{-1}D_2-
(D_2{\cal D}^{-1}D_1)(\mathbb{I}_{N^2-1}-D_1{\cal D}^{-1}D_1)(D_1^{-1}{\cal D}D_2^{-1})(\mathbb{I}_{N^2-1}-D_2{\cal D}^{-1}D_2)
\Bigr)
\nonumber\\
&=&
{\rm det}\Bigl(D_2{\cal D}^{-1}(D_1)^2{\cal D}^{-1}D_2
 -\mathbb{I}_{N^2-1} +D_2{\cal D}^{-1}D_2 +
D_2{\cal D}^{-1}(D_1)^2 D_2^{-1} 
-D_2{\cal D}^{-1}(D_1)^2{\cal D}^{-1}D_2
\Bigr)\nonumber\\
&=&
(-1)^{N^2-1}~
{\rm det}(D_2^{-1}D_2)
{\rm det}\Bigl( \mathbb{I}_{N^2-1}- D_2{\cal D}^{-1}D_2 - D_2{\cal D}^{-1}(D_1)^2 D_2^{-1}  \Bigr)\nonumber\\
&=&(-1)^{N^2-1}~{\rm det}
\Bigl(D_2^{-1} \bigl( 
\mathbb{I}_{N^2-1} -D_2{\cal D}^{-1}D_2 -D_2{\cal D}^{-1}(D_1)^2 D_2^{-1} \bigr)D_2\Bigr)
\nonumber\\
&=&(-1)^{N^2-1}~{\rm det}({\cal D}^{-1}(\del_3)^2),
\label{eq:relationaxialdet}
\end{eqnarray}
where we have used ${\cal D}=(D_1)^2 +(D_2)^2 +(\del_3)^2$ in the last line in Eq.(\ref{eq:relationaxialdet}). Hence, we obtain that
\begin{equation}
{\rm det}(M)={\rm det}\bigl({\cal D}^{-1}(\del_3)^2\bigr),
\end{equation}
which is the equation (\ref{eq:determinantaxial}) in the text.
\subsection{Coulomb gauge}
\label{sec:detcoul}
Instead of evaluating ${\rm det}(N)$ directly, let us consider the determinant defined by 
\begin{equation}
{\rm det}\Biggl(
\begin{pmatrix} 
0 & \mathbb{I}_{N^2-1} \\
\mathbb{I}_{N^2-1} & 0\\
\end{pmatrix} N'
\Biggr),
\label{eq:nprimerel}
\end{equation}
where $N'$ is given by Eq.(\ref{eq:nprimeexp}) and is written as
\begin{equation}
(N^{\prime})_{kl}^{ab}=\begin{pmatrix}
\mathbb{I}_{N^2-1} - {\tilde C}_1^{\prime}{\cal D}^{-1}C_1 & -{\tilde C}^{\prime}_1{\cal D}^{-1}C_2 \\
 -{\tilde C}_2^{\prime} {\cal D}^{-1}C_1& \mathbb{I}_{N^2-1}- {\tilde C}_2^{\prime}{\cal D}^{-1}C_2 \\
\end{pmatrix}.
\end{equation}
The equation (\ref{eq:nprimerel}) becomes, by using the formula (\ref{eq:matformula}),
\begin{eqnarray}
{\rm (\ref{eq:nprimerel})}&=&
{\rm det}\begin{pmatrix}
 -{\tilde C}_2^{\prime} {\cal D}^{-1}C_1 &\mathbb{I}_{N^2-1}- {\tilde C}_2^{\prime}{\cal D}^{-1}C_2 \\
\mathbb{I}_{N^2-1} - {\tilde C}_1^{\prime}{\cal D}^{-1}C_1&  -{\tilde C}^{\prime}_1{\cal D}^{-1}C_2 \\
\end{pmatrix} 
\nonumber\\
&=&
{\rm det}\Bigl(-{\tilde C}_2^{\prime}{\cal D}^{-1}C_1\Bigr)\nonumber\\
&&\times 
{\rm det}\Bigl(-{\tilde C}_1^{\prime}{\cal D}^{-1}C_2 -(\mathbb{I}_{N^2-1} - {\tilde C}_1^{\prime}{\cal D}^{-1}C_1)
(-{\tilde C}_2^{\prime} {\cal D}^{-1}C_1)^{-1}
(\mathbb{I}_{N^2-1}- {\tilde C}_2^{\prime}{\cal D}^{-1}C_2)
\Bigr)
\nonumber\\
&=&{\rm det}\Bigl(-\mathbb{I}_{N^2-1} +{\tilde C}_2^{\prime}{\cal D}^{-1}C_2
+{\tilde C}_2^{\prime}{\cal D}^{-1}C_1 {\tilde C}_1^{\prime}({\tilde C}_2^{\prime})^{-1}\Bigr)\nonumber\\
&=&(-1)^{N^2-1}~{\rm det}({\tilde C}_2^{\prime -1} {\tilde C}_2^{\prime})
{\rm det}\Bigl(\mathbb{I}_{N^2-1} -{\tilde C}_2^{\prime}{\cal D}^{-1}C_2
-{\tilde C}_2^{\prime}{\cal D}^{-1}C_1 {\tilde C}_1^{\prime}({\tilde C}_2^{\prime})^{-1}\Bigr)
\nonumber\\
&=&(-1)^{N^2-1}~
{\rm det}(\mathbb{I}_{N^2-1} - {\cal D}^{-1}C_2{\tilde C}_2^{\prime} -{\cal D}^{-1}C_1{\tilde C}_1^{\prime})
\nonumber\\
&=&(-1)^{N^2-1}~{\rm det}({\cal D}^{-1})
{\rm det}({\cal D} - C_2{\tilde C}_2^{\prime} -C_1{\tilde C}_1^{\prime})
\nonumber\\
&=&
(-1)^{N^2-1}~{\rm det}({\cal D}^{-1})
{\rm det}({\cal D} - C_k{\tilde C}_k^{\prime})\nonumber\\
&=&(-1)^{N^2-1}~{\rm det}({\cal D}^{-1})
{\rm det}({\tilde \Delta})\nonumber\\
&=&
(-1)^{N^2-1}~{\rm det}({\cal D}^{-1}){\rm det}(\Delta^{-1})({\rm det}(\del_iD_i))^2,
\label{eq:detnprime}
\end{eqnarray}
where we have used the equation (\ref{eq:ctildecrel1app})
and $\tilde\Delta=\del\cdot D\Delta^{-1}\del\cdot D$. Hence, we obtain that
\begin{equation}
{\rm det}\Biggl(
\begin{pmatrix} 
0 & \mathbb{I}_{N^2-1} \\
\mathbb{I}_{N^2-1} & 0\\
\end{pmatrix} N'
\Biggr)=(-1)^{N^2-1}~{\rm det}({\cal D}^{-1}){\rm det}(\Delta^{-1})({\rm det}(\del_iD_i))^2 .
\label{eq:nprimerel2}
\end{equation}
On the other hand, it is easy to see that
\begin{equation}
{\rm det}\bigl(\delta_{kl} -\del_k\Delta^{-1}\del_l\bigr)
={\rm det}\begin{pmatrix}
1-\Delta^{-1}(\del_1)^2 & -\Delta^{-1}\del_1\del_2\\
 -\Delta^{-1}\del_2\del_1 & 1-\Delta^{-1}(\del_2)^2\\
\end{pmatrix}={\rm det}(\Delta^{-1}(\del_3)^2),
\end{equation}
so that we have
\begin{eqnarray}
{\rm det}\Biggl(
\begin{pmatrix} 
0 & \mathbb{I}_{N^2-1} \\
\mathbb{I}_{N^2-1} & 0\\
\end{pmatrix} N'
\Biggr)&=&(-1)^{N^2-1}~{\rm det}(N')
\stackrel{{\rm (\ref{eq:nprime})}}{=}(-1)^{N^2-1}~{\rm det}\Big(\bigl(\delta_{kl} -\del_k\Delta^{-1}\del_l\bigr)N\Bigr)\nonumber\\
&=&(-1)^{N^2-1}~{\rm det}(\Delta^{-1}(\del_3)^2){\rm det}(N)
\label{eq:nprimerel3}
\end{eqnarray}
From Eqs.(\ref{eq:nprimerel2}) and (\ref{eq:nprimerel3}), we obtain that
\begin{equation}
{\rm det}(N)=\frac{{\rm det}({\cal D}^{-1})({\rm det}(\del_iD_i))^2}{{\rm det}((\del_3)^2)}
\end{equation}
We have proved the equation (\ref{eq:determinantcoul}) in the text.

\end{document}